\pdfoutput=1
\documentclass[letterpaper,10pt,journal]{IEEEtran}

\usepackage{amsmath,amssymb,amsfonts}
\usepackage{amsthm}
\usepackage{algorithm}
\usepackage{algorithmic}
\usepackage{array}
\usepackage[caption=false,font=normalsize,labelfont=sf,textfont=sf]{subfig}
\usepackage{textcomp}
\usepackage{graphicx}
\usepackage{cite}
\usepackage{url}

\usepackage{array}
\usepackage{amssymb}

\newcolumntype{L}[1]{>{\raggedright\arraybackslash}p{#1}}
\newcolumntype{C}[1]{>{\centering\arraybackslash}p{#1}}

\newtheorem{definition}{Definition}
\newtheorem{theorem}{Theorem}
\newtheorem{lemma}{Lemma}
\newtheorem{proposition}{Proposition}
\newtheorem{corollary}{Corollary}
\newtheorem{assumption}{Assumption}
\newtheorem{remark}{Remark}

\begin{document}

\title{Prediction-Driven Contract Matching with Controlled Overbooking for Cloud--Edge DDoS Scrubbing Markets}

\author{
Yue~Luo,~\IEEEmembership{Member,~IEEE},
Houyi~Qi,~\IEEEmembership{Member,~IEEE},
Minghui~Liwang,~\IEEEmembership{Senior~Member,~IEEE},
Xinlei~Yi,~\IEEEmembership{Senior~Member,~IEEE},
Han~Jiang, and~Xuejiao~Zheng%
\thanks{Y. Luo (lyue@tongji.edu.cn),
H. Qi (houyiqi@tongji.edu.cn),
M. Liwang (minghuiliwang@tongji.edu.cn),
and X. Yi (xinleiyi@tongji.edu.cn) are with
the Shanghai Research Institute for Intelligent
Autonomous Systems, State Key Laboratory of
Autonomous Intelligent Unmanned Systems,
Frontiers Science Center for Intelligent
Autonomous Systems, Tongji University,
Shanghai, China.}%
\thanks{H. Jiang (danielblack@foxmail.com)
is with Pangoal Institution.}%
\thanks{X. Zheng (zhengxuejiao@18.cn)
is with Eastmoney Securities Co., Ltd.}%
}

\maketitle

\begin{center}
\footnotesize\itshape
This work has been submitted to the IEEE for possible publication.
Copyright may be transferred without notice, after which this version
may no longer be accessible.
\end{center}

\begin{abstract}
Cloud--edge distributed denial-of-service (DDoS) scrubbing
requires ex-ante capacity commitments under uncertain
traffic, heterogeneous budgets, resource costs, and
fulfillment risks. We develop a prediction-driven two-stage
contracting mechanism coordinating business entities (BEs),
edge scrubbing nodes (ESs), and a cloud scrubbing center (CS)
through BE--ES (B\&E) main-service and ES--CS (E\&C)
supplementary contracts. In the first offline phase,
Transformer forecasts generate candidate reserved scrubbing
rates, generalized Nash bargaining determines candidate
prices, and interim many-to-one B\&E matching permits
controlled overbooking under scenario-based aggregate ES
shortfall-risk assessment. In the second offline phase,
formal E\&C reservations are established subject to capacity
and payment constraints; their finalized rates and prices
update ES-side proxy scores and guide localized risk-aware
B\&E contract adjustment. During online fulfillment, an
edge-first, cloud-assisted weighted proportional allocation
policy allocates scrubbing capacity under finalized contracts
without rematching, renegotiation, or additional cloud
reservations. Experiments with CICDDoS2019-derived workloads
show that the proposed mechanism satisfies the prescribed
empirical shortfall-risk criterion across all 100 evaluated
horizon-instance pairs, whereas the NoRisk ablation violates it
in all 100. Runtime measurements indicate lightweight online fulfillment
relative to offline contract formation. The evaluated
configurations also exhibit limited overall scrubbing coverage
and near-zero realized cloud utilization.
\end{abstract}


\section{Introduction}
\label{sec:introduction}

\IEEEPARstart{T}{he} growing reliance of online services on
distributed computing and networked infrastructures makes
service availability a critical operational requirement.
Distributed denial-of-service (DDoS) attacks can exhaust
communication and processing resources, disrupting service
continuity~\cite{zargar2013survey,yan2016sdn}.
Cloud--edge collaborative DDoS mitigation combines local
edge processing with supplementary cloud
assistance~\cite{li2021optimal}.
Traffic prediction has also been explored to support
proactive resource coordination in cloud--edge
networks~\cite{chen2020traffic}.

In this setting, scrubbing capacity is not merely a physical
resource to be scheduled, but also an economic resource to be
contracted. Business entities (BEs) compete for limited
protection capacity under budget constraints, edge scrubbing
nodes (ESs) incur operating costs and fulfillment risks, and
a cloud scrubbing center (CS) provides supplementary capacity
subject to physical and economic constraints.
Related research has investigated resource optimization in
wireless networks and learning-based multi-agent coordination
in cloud--edge systems~\cite{graph2025survey,transformer2025cloudedge}.
Existing DDoS studies have explored online traffic diversion~\cite{jiao2019online}, auction-based service procurement~\cite{you2020scheduling}, edge resource allocation~\cite{he2022game}, and predictive cloud--edge coordination~\cite{zhou2024online}. Nevertheless, the coordination of advance edge-side service
commitments and cloud-side supplementary reservations under
uncertain future traffic remains less explored.

Ex-ante contracting in such markets presents three coupled
challenges. First, future suspicious-traffic demand is unknown
at contract formation, requiring imperfect predictions to be
translated into economically feasible commitments while
accounting for prediction errors and contractual risks.
Second, conservative reservations may underutilize scarce
edge capacity, whereas aggressive commitments increase
potential shortfalls and compensation exposure, creating
a utilization--risk tradeoff~\cite{liwang2022unifying}.
Third, the two contractual layers are interdependent:
interim BE--ES commitments determine supplementary cloud
demand, while the resulting cloud reservations affect ES
fulfillment capability and shortfall risk, potentially
changing which BE--ES contracts can be retained.

To address these challenges, we develop a prediction-driven
two-stage contract matching and fulfillment mechanism.
The offline stage comprises two phases. First, traffic
predictions guide candidate BE--ES main-service contract
generation and bilateral bargaining, followed by interim
many-to-one matching under concurrency and controlled
overbooking constraints. Second, interim BE--ES commitments
induce ES--CS supplementary reservation demand.
Candidate ES--CS contracts are priced through bilateral
bargaining and selected subject to cloud capacity and
economic constraints. The resulting cloud reservations
are then incorporated into final risk-aware BE--ES contract
adjustment before demand realization. In the online stage,
finalized contracts are fulfilled through edge-first,
cloud-assisted weighted proportional allocation (WPA)
without contract rematching, price renegotiation, or
additional cloud reservation. 

The main contributions are summarized as follows.

\begin{itemize}

\item
\emph{Ex-ante two-layer DDoS service-market modeling:}
We formulate cloud--edge DDoS scrubbing as a two-layer
contracting problem under uncertain traffic demand,
capturing heterogeneous participant objectives, budgets,
physical capacities, and fulfillment risks. The model
incorporates BE--ES main-service and ES--CS supplementary
contracts, distinguishing offline contract formation
from online fulfillment after demand realization.

\item
\emph{Prediction-driven cross-layer contract formation
and fulfillment:}
We develop a staged mechanism integrating forecast-based
BE--ES contract generation, bilateral bargaining, interim
matching with controlled overbooking, and ES--CS
supplementary reservation. The allocated cloud capacity
feeds back into final risk-aware BE--ES contract adjustment,
forming a cross-layer contract-formation loop.
An edge-first, cloud-assisted WPA policy subsequently
fulfills finalized contracts without online rematching,
renegotiation, or additional cloud reservation.

\item
\emph{Theoretical analysis and trace-driven evaluation:}
We establish finite termination and final contract feasibility,
characterize conditional ex-ante proxy rationality, and derive
conditional restricted final-reapplication stability over the explored
submarket under the stated choice-function assumptions.
For fixed contracts
and reservations, online fulfillment is shown to be
feasible, work-conserving, and Pareto efficient in
fulfilled contractual traffic. Trace-derived experiments
using CICDDoS2019 evaluate contract admission, scrubbing
coverage, resource utilization, social welfare, shortfall
risk, and runtime against benchmark methods and ablation
variants.

\end{itemize}

The remainder of this paper is organized as follows.
Section~\ref{sec:related_work} reviews related work.
Sections~\ref{sec:system_model}--\ref{sec:method}
present the system model, problem formulation, and
proposed mechanism. Section~\ref{sec:performance_evaluation}
reports the experimental results, and
Section~\ref{sec:conclusion} concludes the paper.
Detailed theoretical proofs are provided in the
supplementary material.

\section{Related Work}
\label{sec:related_work}

\subsection{Cloud--Edge DDoS Scrubbing and Proactive Defense}
\label{subsec:rw_ddos}

Prior studies have investigated DDoS mitigation through traffic
diversion, resource scheduling, and service procurement.
Jiao et al.~\cite{jiao2019online} studied online traffic
diversion and cloud scrubbing under uncertain current inputs
and switching costs. You et al.~\cite{you2020scheduling}
developed randomized online auctions for external
DDoS scrubbing-service procurement.
Zhou et al.~\cite{zhou2024online} developed
prediction-oblivious and prediction-aware algorithms
for coordinated cloud--edge scrubbing with local-scrubbing
prioritization. Beyond scheduling,
Deng et al.~\cite{deng2022resource} investigated DDoS-aware
resource provisioning in MEC-enabled SDVNs using GNN-based
collaborative deep reinforcement learning, while
Wu et al.~\cite{wu2026cooperative} studied distributed
cooperative DDoS defense through game-theoretic modeling.
Neira et al.~\cite{neira2023intelligent} investigated DDoS
prediction using early warning signals from attack preparation.

These studies primarily address operational resource allocation,
service procurement, cooperative defense, or attack prediction.
Our work focuses on ex-ante cross-layer scrubbing commitments
under uncertain future traffic demand.

\subsection{Contract and Market Mechanisms for Cloud--Edge Provisioning}
\label{subsec:rw_market}

Qiu et al.~\cite{qiu2022auction} surveyed auction and
mechanism-design approaches for edge computing.
For DDoS mitigation, He et al.~\cite{he2022game}
developed a game-theoretic resource-allocation approach,
while Pan et al.~\cite{pan2025auction} combined a
potential game for request competition with auction-based
strategy adoption across timeslots.
Lee et al.~\cite{lee2024ips} investigated joint resource
allocation and intrusion-prevention-system deployment
through game-theoretic pricing.
Beyond security-specific settings,
You et al.~\cite{you2024incentive} developed a
privacy-preserving two-level auction mechanism for
computation offloading, while
Wang et al.~\cite{wang2020matching} employed stable matching
for joint task offloading and payment determination.
Related mechanism-design studies have investigated
budget-feasible procurement with partial allocations
and linear posted pricing for divisible
resources~\cite{amanatidis2025partial,caragiannis2026divisible}.

Closely related to our work,
Qi et al.~\cite{qi2024bridge} developed a hybrid
futures-and-spot trading framework for cloud-aided mobile
edge networks, incorporating overbooking and risk control
through futures-stage (OA-CLM) and spot-stage (OS-CLM)
cross-layer matching.
While their framework addresses computation-resource trading,
our work considers ex-ante DDoS scrubbing commitments under
uncertain suspicious-traffic demand, where interim BE--ES
contracts induce supplementary ES--CS reservation demand
and the allocated cloud capacity is incorporated into
final risk-aware BE--ES adjustment.

\subsection{Advance Reservation and Overbooking under Uncertainty}
\label{subsec:rw_reservation_overbooking}

Advance reservation and overbooking have been studied under
uncertain resource demand and availability.
Chaisiri et al.~\cite{chaisiri2012optimization} employed
stochastic programming for cloud-resource provisioning under
demand and price uncertainty.
Dazhi et al.~\cite{dazhi2025slicing} investigated
LSTM-assisted traffic prediction for admission control
in non-terrestrial network slicing.
Barbarossa et al.~\cite{barbarossa2017overbooking}
studied radio and computation resource overbooking
to address channel intermittency in mobile edge computing,
while Zhang et al.~\cite{zhang2018dynamic} investigated
dynamic fog-resource overbooking using
resource-utilization prediction.
Sexton et al.~\cite{sexton2020provisioning} considered
forecast-based auxiliary resource provisioning and
overbooking for network slices.
In market-based provisioning,
Sheng et al.~\cite{sheng2020futures} studied futures-based
resource trading using historical supply--demand information.
Liwang et al.~\cite{liwang2022unifying} integrated
overbooking into a hybrid futures-and-spot market,
while Liwang and Wang~\cite{liwang2022overbooking}
investigated overbooking-enabled advance contracts
across user--edge and edge--cloud layers.
Ma et al.~\cite{ma2026following} studied risk-aware
task scheduling with overbooking in edge clouds.

Table~\ref{tab:related_work_comparison} compares
representative studies in terms of application,
decision timing, uncertainty handling, overbooking,
and coordination structure. Our work integrates
prediction-driven DDoS scrubbing contracts with
controlled overbooking and supplementary cloud
reservation, whose outcomes inform final risk-aware
B\&E adjustment before demand realization.

\begin{table*}[!t]
\centering
\caption{Comparison of representative studies on DDoS
scrubbing and resource provisioning.}
\label{tab:related_work_comparison}
\footnotesize
\setlength{\tabcolsep}{3pt}
\renewcommand{\arraystretch}{1.12}
\begin{tabular}{@{}
L{0.12\textwidth}
L{0.12\textwidth}
L{0.16\textwidth}
L{0.195\textwidth}
C{0.072\textwidth}
L{0.255\textwidth}@{}}
\hline
\textbf{Reference} &
\textbf{Application} &
\textbf{Decision timing} &
\textbf{Uncertainty handling} &
\textbf{\shortstack[c]{Over-\\booking}} &
\textbf{Coordination structure} \\
\hline

Jiao et al.~\cite{jiao2019online} &
Cloud DDoS scrubbing &
Online scheduling &
Online optimization with unknown current inputs &
-- &
Traffic diversion and cloud scrubbing \\

You et al.~\cite{you2020scheduling} &
Cloud DDoS scrubbing &
Online auction &
Online auctions without future information &
-- &
ISP--security-provider procurement \\

Zhou et al.~\cite{zhou2024online} &
Cloud--edge DDoS scrubbing &
Predictive online scheduling &
Prediction-aware online scheduling &
-- &
Coordinated edge--cloud scrubbing \\

Sexton et al.~\cite{sexton2020provisioning} &
Network slicing &
Long-term assurance; short-term offers &
Forecast-based probabilistic resource availability &
$\checkmark$ &
Assured and auxiliary slice resources \\

Liwang and Wang~\cite{liwang2022overbooking} &
Edge computing &
Advance contracts &
Historical statistics and risk assessment &
$\checkmark$ &
User--edge and edge--cloud contracts \\

Qi et al.~\cite{qi2024bridge} &
Edge computing &
Futures and online spot matching &
Stochastic participation and contractual risk control &
$\checkmark$ &
Cross-layer futures and spot matching \\

\textbf{Our work} &
\textbf{Cloud--edge DDoS scrubbing} &
\textbf{Ex-ante contracts; online fulfillment} &
\textbf{Causal traffic forecasting and empirical shortfall-risk assessment} &
$\checkmark$ &
\textbf{Interim B\&E matching, E\&C reservation,
and final B\&E adjustment} \\
\hline
\end{tabular}

\vspace{1mm}
\parbox{0.97\textwidth}{\footnotesize
$\checkmark$ denotes an explicit overbooking mechanism;
-- indicates that overbooking is not a principal feature
of the cited method.}
\end{table*}

\section{System Model}
\label{sec:system_model}

\subsection{System Overview}
\label{subsec:system_overview}

We consider a cloud--edge collaborative service market for scrubbing
suspicious traffic associated with DDoS attacks.\footnote{We assume
that upstream security mechanisms have already identified suspicious
traffic~\cite{zargar2013survey,yan2016sdn}. Therefore, this work focuses
on scrubbing-capacity reservation, contract formation, and service
fulfillment rather than attack detection or traffic classification.}
This market comprises BEs, ESs, and a centralized CS.
BEs reserve main scrubbing services from ESs through BE--ES
main-service contracts, while ESs may reserve supplementary scrubbing
capacity from the CS through ES--CS supplementary
contracts.\footnote{We consider a setting with a single cloud scrubbing
center. Accordingly, the BE--ES layer is modeled as a many-to-one
contract-matching market, whereas the ES--CS layer is modeled as a
supplementary capacity reservation and allocation problem.
Hence, blocking pairs are defined only for the BE--ES
contract-matching layer, and the stability result established later
for this layer holds under the assumptions stated with the
corresponding theorem.}
The service horizon is partitioned into a set of timeslots
$\mathcal{T}=\{1,2,\ldots,|\mathcal{T}|\}$, where each timeslot
consists of a demand-generation sub-timeslot (DGsT) and a
demand-processing sub-timeslot (DPsT), with durations $\Delta t'$
and $\Delta t$, respectively, and total duration
$\Delta t^{\mathsf{all}}=\Delta t'+\Delta t$.
For each target timeslot, the offline stage is completed before the
corresponding actual suspicious-traffic demand is observed.
Based on predicted demand, the market establishes BE--ES main-service
contracts and, when needed, ES--CS supplementary contracts subject to
controlled edge overbooking, cloud-capacity constraints, and
fulfillment-risk requirements.
After the actual suspicious traffic is realized, the online stage
executes the contracts established offline through an edge-first and
cloud-assisted fulfillment policy during the DPsT, without contract
rematching, price renegotiation, or additional cloud-side reservation.

\subsection{Market Participants and Their Characteristics}
\label{subsec:market_participants}

\paragraph{Business entities.}
Let
$\mathcal{B}
=
\{b_1,b_i,\ldots,b_{|\mathcal{B}|}\}$
denote the set of BEs.
Since different BEs may exhibit heterogeneous suspicious-traffic demands,
reservation budgets, and service requirements, the ex-ante state of
$b_i$ in timeslot $t$ is characterized by
$\left(
\hat{d}_i^{B,(t)},
\overline{C}_i^{B,(t)},
\underline{\mathcal{Q}}_i^{B,\mathsf{cov},(t)}
\right)$.
Here, $\hat{d}_i^{B,(t)}$ denotes the predicted volume of suspicious
traffic destined for $b_i$ in timeslot $t$ (e.g., Gb),
$\overline{C}_i^{B,(t)}$ represents its maximum payment budget for
scrubbing service, and
$\underline{\mathcal{Q}}_i^{B,\mathsf{cov},(t)}$ denotes its minimum required
scrubbing coverage ratio.
The realized suspicious-traffic volume destined for $b_i$ in timeslot
$t$ is denoted by $d_i^{B,(t)}$.

\paragraph{Edge scrubbing nodes.}
Let
$\mathcal{S}
=
\{s_1,s_j,\ldots,s_{|\mathcal{S}|}\}$
denote the set of ESs.
Each $s_j$ in timeslot $t$ is characterized by
$\left(
f_j^S,
N_j^{\mathsf{max}},
c_j^{S,(t)},
\mathcal{U}_j^S
\right)$.
Here, $f_j^S$ denotes the physical scrubbing-rate limit of $s_j$
(e.g., Gbps), $N_j^{\mathsf{max}}$ is the maximum number of BEs that
$s_j$ can serve concurrently within one timeslot, and $c_j^{S,(t)}$
denotes its unit scrubbing cost.
The parameter $\mathcal{U}_j^S\in(0,1]$ characterizes the historical reliability
and fulfillment level of $s_j$ and is used as a reliability factor
during B\&E contract formation.

\paragraph{Cloud scrubbing center.}
The CS in timeslot $t$ is characterized by
$\left(
f^C,
c^{C,(t)}
\right)$,
where $f^C$ denotes its physical scrubbing-rate limit (e.g., Gbps) and
$c^{C,(t)}$ denotes its unit scrubbing cost.

\subsection{Two-Layer Contract Model}
\label{subsec:contract_model}

To support ex-ante scrubbing-capacity reservation, we define two types
of contracts for each timeslot $t$: BE--ES main-service contracts
(B\&E contracts) and ES--CS supplementary contracts (E\&C contracts).

\paragraph{B\&E contracts.}
For each BE--ES pair $(b_i,s_j)$, the B\&E contract for timeslot $t$
is defined as
$
\mathbb{C}_{i,j}^{B\leftrightarrow E,(t)}
=
\left(
x_{i,j}^{B\leftrightarrow E,(t)},
p_{i,j}^{B\leftrightarrow E,(t)},
q_{i,j}^{E\rightarrow B,(t)}
\right).
$
Here, $x_{i,j}^{B\leftrightarrow E,(t)}$ denotes the scrubbing rate
reserved by BE $b_i$ from ES $s_j$ for timeslot $t$ (e.g., Gbps),
$p_{i,j}^{B\leftrightarrow E,(t)}$ denotes the corresponding unit
reservation price, and $q_{i,j}^{E\rightarrow B,(t)}$ denotes the unit
compensation paid by $s_j$ to $b_i$ for unfulfilled contractual
scrubbing demand.
The reservation price is jointly determined by $b_i$ and $s_j$
through the bilateral bargaining mechanism introduced later.
To accommodate prediction errors and asynchronous traffic peaks across
BEs, an ES may employ controlled overbooking of its local scrubbing
capacity.
Let $z_{i,j}^{(t)}\in\{0,1\}$ indicate whether a B\&E contract is
formed between $b_i$ and $s_j$ in timeslot $t$.
The aggregate scrubbing rate booked at $s_j$ is defined as
\(R_{j}^{S,\mathsf{book},(t)}
=\sum_{b_i\in\mathcal{B}}
z_{i,j}^{(t)}
x_{i,j}^{B\leftrightarrow E,(t)}\)
Accordingly, the booking ratio of $s_j$ is defined as
$
\mathsf{OR}_{j}^{S,(t)}
=
\frac{
R_{j}^{S,\mathsf{book},(t)}
}{
f_j^S
}.
\label{eq:edge_overbooking_factor}
$
Controlled overbooking requires
$\mathsf{OR}_{j}^{S,(t)}
\leq
\mathsf{OR}_{j}^{S,\mathsf{max}}$,
where
$\mathsf{OR}_{j}^{S,\mathsf{max}}\geq1$
denotes the maximum controlled-overbooking ratio allowed for $s_j$.

\paragraph{E\&C contracts.}
For each ES $s_j$, the E\&C supplementary contract for timeslot $t$
is defined as
$
\mathbb{C}_{j}^{E\leftrightarrow C,(t)}
=
\left(
x_{j}^{E\leftrightarrow C,(t)},
p_{j}^{E\leftrightarrow C,(t)}
\right).
$
Here, $x_{j}^{E\leftrightarrow C,(t)}$ denotes the supplementary
scrubbing rate reserved by $s_j$ from the CS for timeslot $t$
(e.g., Gbps), and $p_{j}^{E\leftrightarrow C,(t)}$ denotes the
corresponding unit reservation price, jointly determined by $s_j$
and the CS through the bilateral bargaining mechanism introduced
later.
The E\&C contract provides supplementary capacity when the interim
B\&E booked rate of an ES exceeds its local physical scrubbing
capacity.
Unlike the B\&E layer, overbooking is not permitted at the E\&C layer.
The aggregate supplementary scrubbing rate reserved from the CS must
therefore satisfy
\begin{equation}
\sum_{s_j\in\mathcal{S}}
x_{j}^{E\leftrightarrow C,(t)}
\leq
f^C.
\label{eq:cloud_no_overbooking}
\end{equation}
Unused supplementary reservation is treated as underutilized reserved
capacity on the ES side rather than a fulfillment failure of the CS.

\subsection{Utility and Risk Modeling}
\label{subsec:utility_risk}

During the DPsT of timeslot $t$, online fulfillment follows
the finalized B\&E and retained E\&C contracts. For each
finalized B\&E contract $(b_i,s_j)$, let
$y_{i,j}^{B\leftrightarrow E,(t)}$ and
$y_{i,j}^{E\leftrightarrow C,(t)}$ denote the realized
suspicious-traffic volumes scrubbed locally by $s_j$
and by the CS using supplementary capacity reserved for
$s_j$, respectively. These quantities are used to specify
realized utilities, scrubbing coverage, uncovered traffic,
and compensation. We further define ex-ante risk measures
for offline contract formation.

\subsubsection{BE Utility and Reservation Risk}
\label{subsubsec:be_utility_risk}

For each BE $b_i$, the realized fulfillment outcome determines its
scrubbing value, reservation payment, service-quality loss, and
compensation, while prediction uncertainty induces ex-ante reservation
risk during offline contract formation.
All realized traffic volumes below refer to the DPsT of timeslot $t$.

\paragraph{Realized utility.}
The total volume of suspicious traffic processed for BE $b_i$ is
$y_i^{\mathsf{tot},(t)}
=
\sum_{s_j\in\mathcal{S}}
z_{i,j}^{(t)}
\big(
y_{i,j}^{B\leftrightarrow E,(t)}
+
y_{i,j}^{E\leftrightarrow C,(t)}
\big)$.
The realized scrubbing coverage ratio (SCR) is defined as
\(\mathcal{Q}_i^{B,\mathsf{cov},(t)}
=\min\left\{1,
\frac{y_i^{\mathsf{tot},(t)}}{d_i^{B,(t)}}
\right\}\).
When $d_i^{B,(t)}=0$, we set
\(\mathcal{Q}_i^{B,\mathsf{cov},(t)}=1\).
Accordingly, the realized scrubbing value is
$V_i^{B,(t)}
=
\mathcal{V}_i^B
d_i^{B,(t)}
\mathcal{Q}_i^{B,\mathsf{cov},(t)}$,
where $\mathcal{V}_i^B>0$ denotes the value per unit of suspicious traffic
covered by the scrubbing service.
Under the reservation-based service model, BE $b_i$ pays
$P_i^{B,(t)}
=
\sum_{s_j\in\mathcal{S}}
z_{i,j}^{(t)}
p_{i,j}^{B\leftrightarrow E,(t)}
x_{i,j}^{B\leftrightarrow E,(t)}
\Delta t$
for its finalized B\&E reservation.
The uncovered suspicious-traffic volume is
$y_i^{\mathsf{uncov},(t)}
=
[d_i^{B,(t)}-y_i^{\mathsf{tot},(t)}]^+$,
which incurs the service-quality loss
$L_i^{B,(t)}
=
\mathcal{J}_{i,\mathsf{uncov}}^B
y_i^{\mathsf{uncov},(t)}$,
where $[x]^+=\max\{x,0\}$ and
$\mathcal{J}_{i,\mathsf{uncov}}^B>0$ denotes the loss per unit of
uncovered suspicious traffic.
Compensation applies only to unfulfilled \emph{contractual} demand,
rather than to suspicious traffic beyond the reserved B\&E volume:
\begin{equation}
\begin{aligned}
Q_i^{B,(t)}
={}&
\sum_{s_j\in\mathcal{S}}
z_{i,j}^{(t)}
q_{i,j}^{E\rightarrow B,(t)}
\Big[
\min\{
d_i^{B,(t)},
x_{i,j}^{B\leftrightarrow E,(t)}\Delta t
\}
\\[-0.3ex]
&\qquad
-
y_{i,j}^{B\leftrightarrow E,(t)}
-
y_{i,j}^{E\leftrightarrow C,(t)}
\Big]^+ .
\end{aligned}
\label{eq:be_compensation}
\end{equation}
The resulting slot-level realized utility of $b_i$ is
$U_i^{B,(t)}
=
V_i^{B,(t)}
-
P_i^{B,(t)}
-
L_i^{B,(t)}
+
Q_i^{B,(t)}$.

\paragraph{Offline reservation risk.}
Since $d_i^{B,(t)}$ is unavailable during offline contract formation,
prediction errors may cause a mismatch between predicted and realized
suspicious-traffic demands.
The ex-ante reservation risk of $b_i$ is defined as
$
\mathcal{R}_{i}^{B,(t)}
=
\Pr\left(
\left(
d_i^{B,(t)}
-
\hat{d}_i^{B,(t)}
\right)^2
>
\mathcal{K}_i^B
\right),
\label{eq:be_reservation_risk}
$
where $d_i^{B,(t)}$ is treated as a random variable during offline
contract formation and $\mathcal{K}_i^B>0$ denotes the tolerated squared
prediction-error threshold.
\footnote{In the implementation, the probabilities in the BE
reservation-risk and ES aggregate shortfall-risk definitions are
replaced by their empirical estimates computed from the fixed
validation-residual scenarios. All risk-feasibility statements and
experiments in this paper refer to these empirical estimates; they
are not distribution-free guarantees on the underlying true
probabilities.}
Let $\overline{\mathcal{R}}_i^B\in[0,1]$ denote the maximum
reservation-risk probability tolerated by $b_i$.

\subsubsection{ES Utility and Aggregate Shortfall Risk}
\label{subsubsec:es_utility_risk}

For each ES $s_j$, reservation revenue is obtained from the finalized
B\&E contracts with its matched BEs, whereas the associated costs arise
from local scrubbing, cloud-side supplementary reservation, and
compensation for unfulfilled contractual demand.
We first characterize its realized slot-level utility and then define
the aggregate shortfall risk used for ex-ante risk control.

\paragraph{Realized utility.}
The B\&E reservation revenue of $s_j$ is
$I_j^{S,(t)}
=
\sum_{b_i\in\mathcal{B}}
z_{i,j}^{(t)}
p_{i,j}^{B\leftrightarrow E,(t)}
x_{i,j}^{B\leftrightarrow E,(t)}
\Delta t$.
Its realized local scrubbing cost is
$C_j^{S,\mathsf{loc},(t)}
=
c_j^{S,(t)}
\sum_{b_i\in\mathcal{B}}
z_{i,j}^{(t)}
y_{i,j}^{B\leftrightarrow E,(t)}$,
whereas the cost of its retained E\&C supplementary reservation is
$C_j^{S,\mathsf{cld},(t)}
=
p_j^{E\leftrightarrow C,(t)}
x_j^{E\leftrightarrow C,(t)}
\Delta t$.
The total compensation paid by $s_j$ to its matched BEs
for unfulfilled contractual demand is
$P_j^{S,(t)}
=\sum_{b_i\in\mathcal B}z_{i,j}^{(t)}Q_i^{B,(t)}$,
where $Q_i^{B,(t)}$ is defined in
\eqref{eq:be_compensation}.
Accordingly, the slot-level realized utility of $s_j$ is
$U_j^{S,(t)}
=I_j^{S,(t)}
-C_j^{S,\mathsf{loc},(t)}
-C_j^{S,\mathsf{cld},(t)}
-P_j^{S,(t)}$.

\paragraph{Aggregate shortfall risk.}
Controlled B\&E overbooking allows the aggregate reserved rate associated
with $s_j$ to exceed its local physical scrubbing-rate limit.
Although supplementary cloud capacity can mitigate the resulting
capacity pressure, high-demand realizations may still cause the aggregate
contractual demand to exceed the combined edge and cloud capacities
available during the DPsT.

For a candidate BE set
$\mathcal{B}_j^{(t)}
=
\{\,b_i\in\mathcal{B}\mid z_{i,j}^{(t)}=1\,\}$,
the aggregate contractual demand is defined as
$\widetilde{H}_j^{(t)}(\mathcal{B}_j^{(t)})
=
\sum_{b_i\in\mathcal{B}_j^{(t)}}
\min\{
d_i^{B,(t)},
x_{i,j}^{B\leftrightarrow E,(t)}\Delta t
\}$.
Given the supplementary scrubbing rate
$x_j^{E\leftrightarrow C,(t)}$ reserved from the CS, the total
scrubbing capacity available during the DPsT is
$(f_j^S+x_j^{E\leftrightarrow C,(t)})\Delta t$.
Since the realized traffic demands are uncertain during offline
contract formation, we define the aggregate shortfall risk as
\(
\mathcal{R}_{j,\mathsf{sh}}^{S,(t)}
\left(\mathcal{B}_j^{(t)},x_j^{E\leftrightarrow C,(t)}\right)
=
\Pr\Bigg(
\Big[
\big(
\widetilde{H}_j^{(t)}(\mathcal{B}_j^{(t)})
-
(f_j^S+x_j^{E\leftrightarrow C,(t)})\Delta t
\big)^+
\Big]^2
>
\mathcal{K}_j^S
\Bigg)
\)
Here, $\mathcal{K}_j^S>0$ denotes the squared aggregate-shortfall tolerance
threshold of $s_j$.

\subsubsection{CS Utility}
\label{subsubsec:cs_utility}

As the sole provider of supplementary scrubbing capacity, the CS
receives reservation revenue from ESs through retained E\&C contracts
and incurs operating costs for the supplementary traffic processed
during the DPsT.
Since E\&C overbooking is not permitted and the aggregate supplementary
reservation is constrained by the physical scrubbing-rate limit of the
CS, no reservation-induced fulfillment risk is modeled on the CS side.

\paragraph{Realized utility.}
The reservation revenue of the CS is
$I^{C,(t)}
=
\sum_{s_j\in\mathcal{S}}
p_j^{E\leftrightarrow C,(t)}
x_j^{E\leftrightarrow C,(t)}
\Delta t$.
Its realized operating cost is
$C^{C,\mathsf{run},(t)}
=
c^{C,(t)}
\sum_{s_j\in\mathcal{S}}
\sum_{b_i\in\mathcal{B}}
z_{i,j}^{(t)}
y_{i,j}^{E\leftrightarrow C,(t)}$.
Accordingly, the slot-level realized utility of the CS is
$U^{C,(t)}
=
I^{C,(t)}
-
C^{C,\mathsf{run},(t)}$.

\section{Core Modeling and Problem Formulation}
\label{sec:core_modeling}

\subsection{Matching and Feasibility Definitions}
\label{subsec:contract_matching_definitions}

Following the two-layer contract structure in
Section~\ref{subsec:contract_model}, we model the B\&E layer as a
many-to-one contract-matching market and the E\&C layer as a
single-CS supplementary-capacity reservation problem. Since each
BE--ES pair has at most one candidate B\&E contract per timeslot,
a BE's preference over candidate contracts can equivalently be
represented by its preference over the corresponding ESs. Let
$\succ_i^B$ and $\succeq_i^B$ denote the strict and weak preferences
of BE $b_i$ over $\mathcal{S}\cup\{\emptyset\}$, where $\emptyset$
denotes the unmatched option, and let $\succ_j^S$ and $\succeq_j^S$
denote those of ES $s_j$ over admissible BE acceptance sets. These
preferences are induced by the phase-specific ex-ante proxy objectives
defined in Section~\ref{sec:method}.

\begin{definition}[Many-to-one B\&E contract matching]
\label{def:be_es_matching}
In timeslot $t$, a B\&E matching is a mapping
$\mathcal{M}^{B\rightarrow E,(t)}:\mathcal{B}\rightarrow\mathcal{S}\cup\{\emptyset\}$,
where $\mathcal{M}^{B\rightarrow E,(t)}(b_i)=s_j$ indicates that $b_i$ is
matched with $s_j$ through a B\&E contract (i.e., $z_{i,j}^{(t)}=1$),
and $\mathcal{M}^{B\rightarrow E,(t)}(b_i)=\emptyset$ indicates that $b_i$
remains unmatched. The set of BEs matched to $s_j$ is
$\mathcal{M}^{E\leftarrow B,(t)}(s_j)
=
\{b_i\in\mathcal{B}\mid\mathcal{M}^{B\rightarrow E,(t)}(b_i)=s_j\}$.
\end{definition}

\begin{definition}[E\&C supplementary-contract assignment]
\label{def:ec_contract_assignment}
In timeslot $t$, the E\&C supplementary-contract assignment is the
set of formed E\&C contracts
$\mathcal{M}^{C,(t)}
=
\left\{
\mathbb{C}_{j}^{E\leftrightarrow C,(t)}
\mid
s_j\in\mathcal{S},
\;
x_j^{E\leftrightarrow C,(t)}>0
\right\}$.
The assignment must satisfy the CS capacity constraint in
\eqref{eq:cloud_no_overbooking}.
\end{definition}

\begin{definition}[Two levels of ES-side feasibility]
\label{def:two_levels_of_es_feasibility}
For $s_j\in\mathcal{S}$ and
$\mathcal{B}_j^{(t)}\subseteq\mathcal{B}$,
$\mathcal{B}_j^{(t)}$ is \emph{interim overbooking-feasible} for $s_j$ if
\begin{equation}
\left|\mathcal{B}_j^{(t)}\right|
\leq N_j^{\mathsf{max}},
\sum_{b_i\in\mathcal{B}_j^{(t)}}
x_{i,j}^{B\leftrightarrow E,(t)}
\leq
\mathsf{OR}_j^{S,\mathsf{max}}f_j^S.
\label{eq:interim_overbooking_feasibility}
\end{equation}
After $x_j^{E\leftrightarrow C,(t)}$ is determined,
$\mathcal{B}_j^{(t)}$ is \emph{final risk-feasible} for $s_j$ if it is
interim overbooking-feasible and additionally satisfies
\begin{equation}
\mathcal{R}_{j,\mathsf{sh}}^{S,(t)}
\left(
\mathcal{B}_j^{(t)},
x_j^{E\leftrightarrow C,(t)}
\right)
\leq
\mathcal{Z}_j^S.
\label{eq:final-risk-aware-feasibility}
\end{equation}
\end{definition}

\begin{definition}[Ex-ante proxy rationality]
\label{def:ex_ante_ir}
Given a B\&E matching $\mathcal{M}^{B\rightarrow E,(t)}$ and the corresponding
E\&C assignment $\mathcal{M}^{C,(t)}$, the resulting offline contract-formation
outcome is \emph{ex-ante proxy-rational} if the following proxy
nonnegativity and contract-feasibility conditions hold, with the
no-contract proxy value normalized to zero.

\textit{(i) BE side:}
for every matched pair $(b_i,s_j)$, the BE-side proxy objective
$\Pi_{i,j}^{B,(t)}$ defined in \eqref{eq:be_preference_score} satisfies \begin{equation}
\Pi_{i,j}^{B,(t)} \geq 0.
\label{eq:ir_be_proxy}
\end{equation}
In addition, the BE-side reservation risk satisfies
\begin{equation}
\mathcal{R}_{i}^{B,(t)} \leq \overline{\mathcal{R}}_{i}^{B}.
\label{eq:ir_be_risk}
\end{equation}

\textit{(ii) ES side:}
for every $s_j\in\mathcal{S}$, the accepted set $\mathcal{B}_j^{(t)}$
satisfies
$\mathcal{R}_{j,\mathsf{sh}}^{S,(t)}(\mathcal{B}_j^{(t)},
x_j^{E\leftrightarrow C,(t)})\leq\mathcal{Z}_j^S$, with $\mathcal{Z}_j^S$ the
maximum tolerated aggregate shortfall-risk probability. If
$\mathcal{B}_j^{(t)}$ is chosen by the final ES-side choice operator,
the updated proxy in \eqref{eq:es_updated_proxy_score} satisfies
\begin{equation}
\Pi_{j}^{S,\mathsf{upd},(t)}
\left(
\mathcal{B}_j^{(t)};
x_j^{E\leftrightarrow C,(t)}
\right)
\geq 0.
\label{eq:ir_es_proxy}
\end{equation}
This proxy condition is not claimed for sets obtained solely by
risk-release. Retained E\&C contracts must also be payment-feasible
in
Section~\ref{subsubsec:es_cs_reservation_and_final_adjustment}.

\textit{(iii) CS side:}
the E\&C assignment satisfies
\eqref{eq:cloud_no_overbooking}, and
\begin{equation}
\sum_{s_j\in\mathcal{S}}
G_{j,k_j^*}^{C,(t)}
\geq 0,
\label{eq:ir_cs_proxy}
\end{equation}
where $G_{j,k}^{C,(t)}$ is defined in
\eqref{eq:cloud_proxy_utility_for_candidate_level}, and $k_j^*$
denotes the candidate level selected for $s_j$ by the CS-side
supplementary-capacity allocation procedure.
\end{definition}

\begin{definition}[B\&E blocking pair]
\label{def:be_es_blocking_pair}
Given a B\&E matching $\mathcal{M}^{B\rightarrow E,(t)}$ and the
supplementary rates
$\{x_j^{E\leftrightarrow C,(t)}\}_{s_j\in\mathcal{S}}$,
a pair $(b_i,s_j)$ with
$\mathcal{M}^{B\rightarrow E,(t)}(b_i)\neq s_j$ is a
\emph{B\&E blocking pair} if
$s_j\succ_i^B\mathcal{M}^{B\rightarrow E,(t)}(b_i)$ and there exists
$\widetilde{\mathcal{B}}_j^{(t)}
\subseteq\mathcal{M}^{E\leftarrow B,(t)}(s_j)$ such that
$
\mathcal{B}_j'^{(t)}
=
\left(
\mathcal{M}^{E\leftarrow B,(t)}(s_j)
\setminus
\widetilde{\mathcal{B}}_j^{(t)}
\right)
\cup\{b_i\}
$
is final risk-feasible and
$\mathcal{B}_j'^{(t)}\succ_j^S
\mathcal{M}^{E\leftarrow B,(t)}(s_j)$.
Here,
$\widetilde{\mathcal{B}}_j^{(t)}=\emptyset$ and
$\widetilde{\mathcal{B}}_j^{(t)}\neq\emptyset$ correspond to
expansion- and replacement-type blocking pairs, respectively.
\end{definition}

\begin{definition}[Restricted final-reapplication stability]
\label{def:final_risk_aware_stability}
Given an E\&C assignment $\mathcal{M}^{C,(t)}$ in timeslot $t$, a B\&E matching
$\mathcal{M}^{B\rightarrow E,(t)}$ is \emph{restricted final-reapplication stable}
if the resulting offline outcome is ex-ante proxy-rational and admits
no B\&E blocking pair within the explored final-reapplication submarket,
under the substitutability assumption in the supplementary material.
\end{definition}

\subsection{Problem Formulation}
\label{subsec:problem_formulation}

For each target timeslot $t$, offline contract formation
determines the final B\&E matching and E\&C supplementary
reservations before demand realization. Since the realized
utilities $U_i^{B,(t)}$, $U_j^{S,(t)}$, and $U^{C,(t)}$
depend on both traffic realization and online fulfillment
outcomes, participants instead use phase-specific ex-ante
proxy objectives: $\Pi_{i,j}^{B,(t)}$ in \eqref{eq:be_preference_score},
$\Pi_j^{S,\mathsf{upd},(t)}$ in
\eqref{eq:es_updated_proxy_score}, and
$G_{j,k}^{C,(t)}$ in
\eqref{eq:cloud_proxy_utility_for_candidate_level}.
The corresponding optimization problems are formulated
below.

\paragraph{BE-side objective.}
Each BE $b_i$ selects at most one ES as its main service provider
while respecting its payment budget and reservation-risk tolerance:
\begin{align}
\mathcal{P}_i^B:\quad
\max_{\mathcal{M}^{B\rightarrow E,(t)}(b_i)}
\quad&
\Pi_{i,j}^{B,(t)},
\quad
\forall b_i\in\mathcal{B},
\label{eq:be_objective}
\\
\mathrm{s.t.}\quad
&
\eqref{eq:ir_be_risk},
\nonumber
\\
&
\mathcal{M}^{B\rightarrow E,(t)}(b_i)\in\mathcal{S}\cup\{\emptyset\},
\quad
\forall b_i\in\mathcal{B},
\tag{\ref{eq:be_objective}a}\label{eq:be_matching_domain}
\\
&
\sum_{s_j\in\mathcal{S}}z_{i,j}^{(t)}\leq 1,
\quad
\forall b_i\in\mathcal{B},
\tag{\ref{eq:be_objective}b}\label{eq:be_single_matching}
\\
&
P_i^{B,(t)}\leq\overline{C}_i^{B,(t)},
\quad
\forall b_i\in\mathcal{B},
\tag{\ref{eq:be_objective}c}\label{eq:be_budget_constraint}
\end{align}
Constraint~\eqref{eq:ir_be_risk} enforces the BE-side
reservation-risk requirement, constraint~\eqref{eq:be_matching_domain}
specifies the feasible matching domain,
constraint~\eqref{eq:be_single_matching} ensures that each BE is
matched with at most one ES, and constraint~\eqref{eq:be_budget_constraint}
imposes the payment budget limit.

\paragraph{ES-side objective.}
Given $x_j^{E\leftrightarrow C,(t)}$, each ES $s_j$ selects its final
accepted BE set to maximize the updated ex-ante proxy objective:
\begin{align}
\mathcal{P}_j^S:\quad
\max_{\mathcal{B}_j^{(t)}\subseteq\mathcal{B}}
\quad&
\Pi_j^{S,\mathsf{upd},(t)}
\left(
\mathcal{B}_j^{(t)};
x_j^{E\leftrightarrow C,(t)}
\right),
\quad
\forall s_j\in\mathcal{S},
\label{eq:es_objective}
\\
\mathrm{s.t.}\quad
&
\eqref{eq:interim_overbooking_feasibility},
\eqref{eq:final-risk-aware-feasibility}.
\nonumber
\end{align}
Constraint~\eqref{eq:interim_overbooking_feasibility} limits the
number of BEs accessing $s_j$ and bounds the total reserved scrubbing
rate by controlled overbooking, while
constraint~\eqref{eq:final-risk-aware-feasibility} limits the
aggregate shortfall risk.

\paragraph{CS-side objective.}
The CS selects one candidate supplementary-reservation level for each
ES to maximize its aggregate ex-ante proxy objective subject to its
physical scrubbing-rate limit:
\begin{align}
\mathcal{P}^C:\quad
\max_{\{k_j\}_{s_j\in\mathcal{S}}}
\quad&
\sum_{s_j\in\mathcal{S}}
G_{j,k_j}^{C,(t)}
\label{eq:cs_objective}
\\
\mathrm{s.t.}\quad
&
\sum_{s_j\in\mathcal{S}}
a_{j,k_j}
\leq
f^C,
\tag{\ref{eq:cs_objective}a}\label{eq:cs_capacity}
\\
&
k_j\in\{0,\ldots,K_j\},
\quad
\forall s_j\in\mathcal{S}.
\tag{\ref{eq:cs_objective}b}\label{eq:cs_candidate_level}
\end{align}
Here, $a_{j,k}$ and $G_{j,k}^{C,(t)}$ denote the candidate
supplementary-reservation level and its corresponding CS-side ex-ante
proxy value, respectively, with their detailed constructions given in
Section~\ref{sec:method}. The optimal level $k_j^*$ determines
$x_j^{E\leftrightarrow C,(t)}=a_{j,k_j^*}$.
Constraint~\eqref{eq:cs_capacity} enforces the CS physical
scrubbing-rate limit, while constraint~\eqref{eq:cs_candidate_level}
restricts the selected level for each ES to its admissible candidate
set.

These three subproblems are tightly coupled: interim B\&E matching
determines supplementary E\&C reservation demand, while
the resulting cloud reservations affect ES effective
fulfillment capacity and shortfall risk, potentially
necessitating final B\&E adjustment. Section~\ref{sec:method}
develops a staged mechanism to coordinate these decisions.

\section{Method Design}
\label{sec:method}

\subsection{Overview}
\label{subsec:mechanism_overview}

\begin{figure*}[!t]
    \centering
    \includegraphics[width=\textwidth]{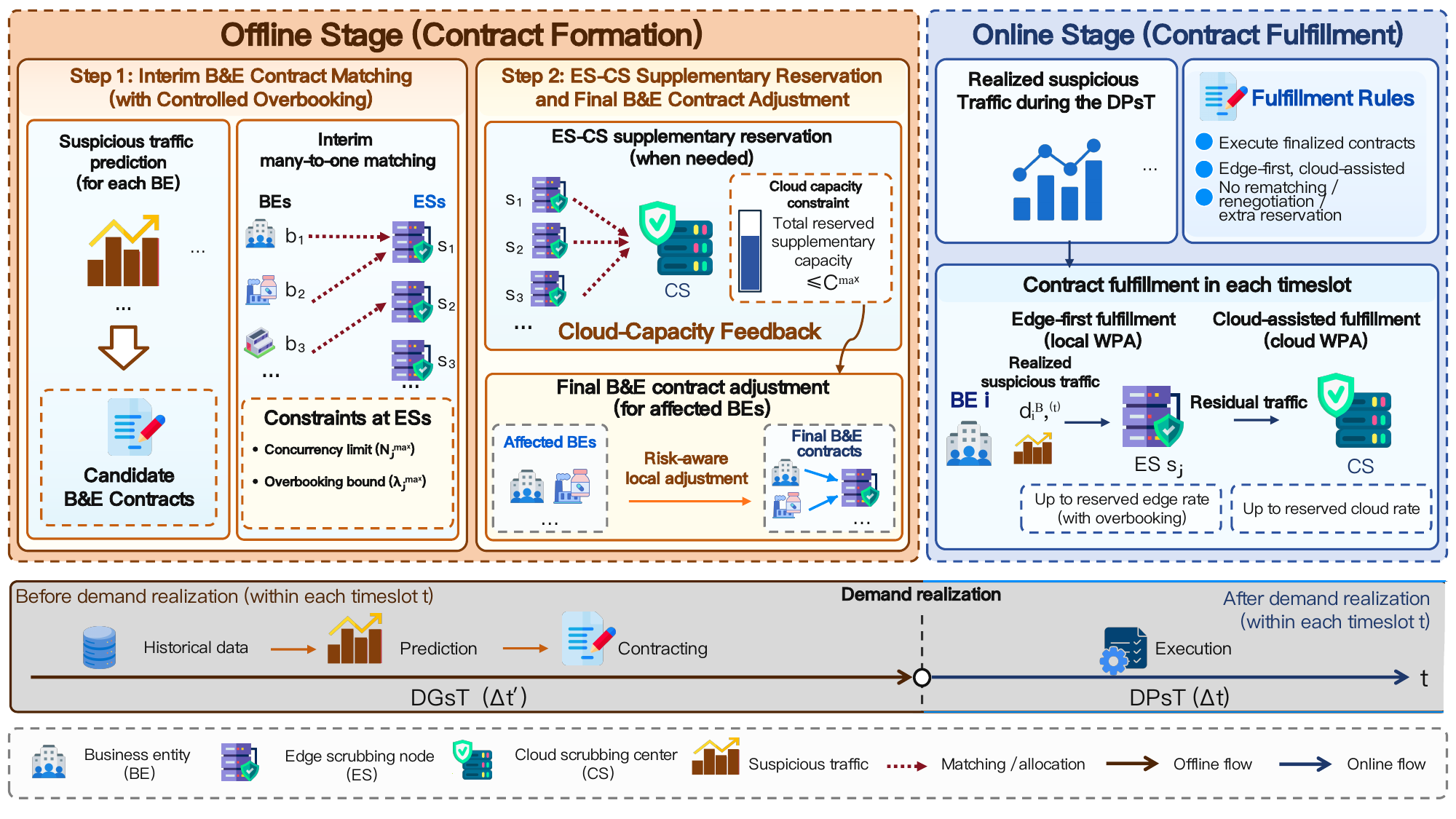}
    \caption{Two-phase offline contract formation and online fulfillment.}
    \label{fig:mechanism_workflow2_lwmh}
\end{figure*}

To coordinate self-interested participants in the cloud--edge DDoS
scrubbing market, we propose a prediction-driven two-stage contract
matching and fulfillment mechanism. In the offline stage,
predicted demand guides candidate B\&E contract generation, with
prices determined via generalized Nash bargaining, followed by
interim matching under controlled overbooking, E\&C supplementary
reservation, and final risk-aware B\&E adjustment. In the online
stage, finalized contracts are fulfilled through an edge-first,
cloud-assisted WPA policy, without contract rematching, price
renegotiation, or additional cloud reservation. The proposed
mechanism is illustrated in Fig.~\ref{fig:mechanism_workflow2_lwmh}.

\subsection{Offline Contract Formation}
\label{subsec:offline_stage}

For each target timeslot, offline contract formation precedes
demand realization and comprises two coupled phases.
Interim B\&E matching determines supplementary E\&C reservation
demand, while the formally allocated cloud capacity is
incorporated into final risk-aware B\&E contract adjustment.

\subsubsection{Prediction-Driven Interim B\&E Contract Matching}
\label{subsubsec:interim_be_matching}

During the first offline phase, predicted suspicious-traffic
demand is converted into a baseline reserved scrubbing rate
for each BE. Under the single-rate reservation design, this
rate is used to generate candidate B\&E main-service contracts
with different ESs. The resulting matching remains interim until formal
E\&C reservations and final risk-aware B\&E contract
adjustment are completed.

\textbf{Step 1: Traffic prediction and baseline rate generation.}
We employ a lightweight Transformer-based predictor
~\cite{vaswani2017attention} to forecast the
suspicious-traffic demand of each BE before candidate B\&E contract
generation.
For BE $b_i$, let
$\mathbf{h}_{i}^{(t)}
=
[d_{i}^{B,(t-L)},d_{i}^{B,(t-L+1)},\ldots,d_{i}^{B,(t-1)}]$
denote the historical suspicious-traffic sequence of length $L$.
The model parameters $\mathcal{E}_i$ are trained by minimizing the mean squared
prediction error over the training set
$\mathcal{T}_{i}^{\mathsf{train}}$, i.e.,
$\mathcal{E}_{i}^{*}
=
\arg\min_{\mathcal{E}_i}
\frac{1}{|\mathcal{T}_{i}^{\mathsf{train}}|}
\sum_{n\in\mathcal{T}_{i}^{\mathsf{train}}}
\big(
F_{\mathcal{E}_i}(\mathbf{h}_{i}^{(n)})
-
d_{i}^{B,(n)}
\big)^2$,
$\forall b_i\in\mathcal{B}$.
The point prediction for the target timeslot is
$\hat{d}_{i}^{B,(t)}
=
[F_{\mathcal{E}_i^*}(\mathbf{h}_{i}^{(t)})]^+$.
Finally, the predicted demand is converted into the baseline candidate
reserved scrubbing rate
$\widetilde{x}_{i}^{B,(t)}
=
\hat{d}_{i}^{B,(t)}/\Delta t$.

\textbf{Step 2: Candidate B\&E contract generation and bilateral pricing.}
For each candidate B\&E pair $(b_i,s_j)$, a candidate B\&E
main-service contract is constructed based on the baseline reserved
scrubbing rate $\widetilde{x}_{i}^{B,(t)}$, i.e.,
$x_{i,j}^{B\leftrightarrow E,(t)}
=
\widetilde{x}_{i}^{B,(t)}$.
Under the single-rate reservation design, each BE uses the same
reservation rate across all candidate ESs. Accordingly, its booked
scrubbing volume during the DPsT is
$y_i^{B,\mathsf{book},(t)}
=
x_{i,j}^{B\leftrightarrow E,(t)}\Delta t$.
A candidate B\&E contract is retained only if it satisfies both the
BE-side reservation-risk requirement and the price-feasibility
condition. For candidate pair $(b_i,s_j)$, the prediction-based
valuation of BE $b_i$ is
$V_{i,j}^{B,\mathsf{pred},(t)}
=
\mathcal{V}_i^B\mathcal{U}_j^S\hat d_i^{B,(t)}$,
while the reservation-cost benchmark of ES $s_j$ is
$C_{i,j}^{S,(t)}
=
c_j^{S,(t)}y_i^{B,\mathsf{book},(t)}$,
where $\mathcal{U}_j^S\in(0,1]$ characterizes the historical service
reliability of ES $s_j$.
For a unit reservation price $p$, the bilateral surpluses are
$S_{i,j}^{B,(t)}(p)
=
V_{i,j}^{B,\mathsf{pred},(t)}
-
p\,y_i^{B,\mathsf{book},(t)}$
for the BE and
$S_{i,j}^{S,(t)}(p)
=
p\,y_i^{B,\mathsf{book},(t)}
-
C_{i,j}^{S,(t)}$
for the ES.
To ensure nonnegative bilateral surpluses while respecting the BE
reservation budget, the feasible unit-price interval is
\(
\underline p_{i,j}^{B\leftrightarrow E,(t)}
=
c_j^{S,(t)},
\overline p_{i,j}^{B\leftrightarrow E,(t)}
=
\min\left\{
\frac{V_{i,j}^{B,\mathsf{pred},(t)}}
{y_i^{B,\mathsf{book},(t)}},
\frac{\overline C_i^{B,(t)}}
{y_i^{B,\mathsf{book},(t)}}
\right\}
\)
The candidate pair is price-feasible only if
$\underline p_{i,j}^{B\leftrightarrow E,(t)}
\leq
\overline p_{i,j}^{B\leftrightarrow E,(t)}$.
For each price-feasible pair, the unit reservation price is determined
through generalized Nash bargaining~\cite{harsanyi1972generalized}:
\begin{equation}
\begin{aligned}
\max_{p}\quad&
\left[
S_{i,j}^{B,(t)}(p)
\right]^{\mathcal{W}_{i,j}^{B}}
\left[
S_{i,j}^{S,(t)}(p)
\right]^{1-\mathcal{W}_{i,j}^{B}}
\\
\mathrm{s.t.}\quad&
\underline p_{i,j}^{B\leftrightarrow E,(t)}
\leq p \leq
\overline p_{i,j}^{B\leftrightarrow E,(t)},
\end{aligned}
\label{eq:be_es_nash_bargaining}
\end{equation}
where $\mathcal{W}_{i,j}^{B}\in(0,1)$ denotes the bargaining weight of
BE $b_i$.
The unconstrained Nash bargaining solution is
$p_{i,j}^{B\leftrightarrow E,\mathsf{NB},(t)}
=
(1-\mathcal{W}_{i,j}^{B})
V_{i,j}^{B,\mathsf{pred},(t)}
/
y_i^{B,\mathsf{book},(t)}
+
\mathcal{W}_{i,j}^{B}c_j^{S,(t)}$.
Accounting for the BE reservation budget, the resulting unit reservation
price is
$
p_{i,j}^{B\leftrightarrow E,*,(t)}
=
\min
\left\{
p_{i,j}^{B\leftrightarrow E,\mathsf{NB},(t)},
\frac{\overline C_i^{B,(t)}}
{y_i^{B,\mathsf{book},(t)}}
\right\}.
\label{eq:be_es_contract_price}
$
The unit compensation for unfulfilled contractual scrubbing demand is
set proportionally as
$q_{i,j}^{E\rightarrow B,(t)}
=
\mathsf{r}^{E\rightarrow B}
p_{i,j}^{B\leftrightarrow E,*,(t)}$,
where $0<\mathsf{r}^{E\rightarrow B}<1$.
The resulting candidate B\&E main-service contract is
$\mathbb C_{i,j}^{B\leftrightarrow E,(t)}
=
(x_{i,j}^{B\leftrightarrow E,(t)},
p_{i,j}^{B\leftrightarrow E,*,(t)},
q_{i,j}^{E\rightarrow B,(t)})$.

\textbf{Step 3: BE preference formation and application.}
For each BE $b_i$, the feasible ES set is defined as
$
\mathcal{S}_{i}^{B,\mathsf{feas},(t)}
=
\left\{
s_j\in\mathcal{S}
\ \middle|\
\mathcal{R}_{i}^{B,(t)}
\leq
\overline{\mathcal{R}}_{i}^{B},
\;
\underline p_{i,j}^{B\leftrightarrow E,(t)}
\leq
\overline p_{i,j}^{B\leftrightarrow E,(t)}
\right\}.
\label{eq:be_feasible_es_set}
$
For each feasible ES
$s_j\in\mathcal{S}_{i}^{B,\mathsf{feas},(t)}$,
the BE-side ex-ante proxy score is
\begin{equation}
\Pi_{i,j}^{B,(t)}
=
\mathcal{V}_i^B
\mathcal{U}_j^S
\hat d_i^{B,(t)}
-
p_{i,j}^{B\leftrightarrow E,*,(t)}
x_{i,j}^{B\leftrightarrow E,(t)}
\Delta t.
\label{eq:be_preference_score}
\end{equation}
The first term represents the prediction-based scrubbing benefit
perceived by BE $b_i$, adjusted by the historical service reliability
of ES $s_j$, whereas the second term is the corresponding reservation
payment.
Based on this proxy score, BE $b_i$ prefers $s_j$ to $s_{j'}$ if
$s_j\succ_i^B s_{j'}$
whenever
$\Pi_{i,j}^{B,(t)}>\Pi_{i,j'}^{B,(t)}$,
with deterministic tie breaking.
The initial preference list is
$\mathcal L_i^{(t),0}
=
\{s_j\in\mathcal S_i^{B,\mathsf{feas},(t)}
\mid
\Pi_{i,j}^{B,(t)}\geq0\}$,
ordered according to $\succ_i^B$.
The working preference list is initialized as
$\mathcal L_i^{(t)}\leftarrow\mathcal L_i^{(t),0}$.
Motivated by the matching-with-contracts framework
~\cite{hatfield2005matching}, each unmatched BE applies to its
most-preferred ES remaining in $\mathcal L_i^{(t)}$.
Once rejected by an ES, the BE removes that ES from its working list
and, if the list remains nonempty, applies to its next preferred ES.

\textbf{Step 4: ES-side interim selection.}
For each ES $s_j$, let
$\mathcal B_j^{\mathsf{cand},(t)}$
denote the union of the newly applying BEs and those tentatively retained
from the previous iteration. ES $s_j$ selects an interim acceptance set
$\mathcal B_j^{(t)}
\subseteq
\mathcal B_j^{\mathsf{cand},(t)}$
subject to its concurrency limit and maximum controlled-overbooking
ratio. For any candidate acceptance set $\mathcal B_j^{(t)}$, the
aggregate reserved scrubbing rate is
$R_j^{S,\mathsf{book},(t)}(\mathcal B_j^{(t)})
=
\sum_{b_i\in\mathcal B_j^{(t)}}
x_{i,j}^{B\leftrightarrow E,(t)}$.
Since the formal E\&C supplementary contract has not yet been
established, let
$\widehat x_j^{E\leftrightarrow C,\mathsf{ref},(t)}$
and
$\widehat p_j^{E\leftrightarrow C,\mathsf{ref},(t)}$
denote a nonbinding reference supplementary reserved rate and unit
reservation price, respectively. These reference values may be obtained
from historical E\&C contract records or specified by system priors and
are used only for interim ES-side evaluation. The formal E\&C
supplementary reserved rate and reservation price are determined in the
subsequent phase.
Using the reference supplementary rate, the aggregate shortfall risk of
$\mathcal B_j^{(t)}$ is evaluated as
$\mathcal R_{j,\mathsf{sh}}^{S,(t)}
(\mathcal B_j^{(t)},
\widehat x_j^{E\leftrightarrow C,\mathsf{ref},(t)})$.
The interim ES-side risk-aware proxy score is defined as
\begingroup
\setlength{\abovedisplayskip}{4pt}
\setlength{\abovedisplayshortskip}{2pt}
\setlength{\belowdisplayskip}{4pt}
\setlength{\belowdisplayshortskip}{2pt}
\begin{equation}
\begin{aligned}
\Pi_j^{S,\mathsf{int},(t)}
\left(
\mathcal B_j^{(t)}
\right)
={}&
\sum_{b_i\in\mathcal B_j^{(t)}}
p_{i,j}^{B\leftrightarrow E,*,(t)}
x_{i,j}^{B\leftrightarrow E,(t)}
\Delta t
\\
&-
c_j^{S,(t)}
\min
\left\{
R_j^{S,\mathsf{book},(t)}
\left(
\mathcal B_j^{(t)}
\right),
f_j^S
\right\}
\Delta t
\\
&-
\widehat p_j^{E\leftrightarrow C,\mathsf{ref},(t)}
\widehat x_j^{E\leftrightarrow C,\mathsf{ref},(t)}
\Delta t
\\
&-
\mathcal{Y}_j^S
\mathcal R_{j,\mathsf{sh}}^{S,(t)}
\left(
\mathcal B_j^{(t)},
\widehat x_j^{E\leftrightarrow C,\mathsf{ref},(t)}
\right),
\end{aligned}
\label{eq:es_interim_proxy_score}
\end{equation}
\endgroup
where $\mathcal{Y}_j^S\geq0$ denotes the weight assigned to aggregate
shortfall risk. The four terms respectively represent the B\&E
reservation revenue, the estimated local scrubbing cost, the reference
E\&C supplementary-reservation payment, and the aggregate shortfall-risk
penalty.
ES $s_j$ determines its interim acceptance set by solving
\begin{align}
\max_{
\mathcal B_j^{(t)}
\subseteq
\mathcal B_j^{\mathsf{cand},(t)}
}
\quad&
\Pi_j^{S,\mathsf{int},(t)}
\left(
\mathcal B_j^{(t)}
\right)
\label{eq:es_interim_choice_objective}
\\
\mathrm{s.t.}\quad&
\eqref{eq:interim_overbooking_feasibility}.
\notag
\end{align}
The resulting ES-side choice is denoted by
$\operatorname{Ch}_j^{S,\mathsf{int},(t)}
(\mathcal B_j^{\mathsf{cand},(t)})
=
\mathcal B_j^{(t),*}$.
The above subset-selection problem is combinatorial and is approximately
solved via dynamic programming. Specifically, the aggregate booked
scrubbing rate is discretized, and the risk-related terms are evaluated
over the resulting discretized states while enforcing the concurrency
constraint. ES $s_j$ tentatively retains the BEs in
$\mathcal B_j^{(t),*}$ and rejects the remaining applicants. Each
rejected BE removes $s_j$ from its current working preference list and,
if the list remains nonempty, applies to its next preferred ES.

\textbf{Step 5: Interim matching outcome.}
The interim B\&E application--rejection process terminates when no BE
remains unmatched while still having a nonempty working preference list.
Based on the ES-selected tentative acceptance sets, define
$z_{i,j}^{\mathsf{int},(t)}=1$ if
$b_i\in\mathcal B_j^{(t),*}$ and
$z_{i,j}^{\mathsf{int},(t)}=0$ otherwise.
Accordingly, the set of BEs matched to ES $s_j$ is
$\mathcal{M}^{E\leftarrow B,\mathsf{int},(t)}(s_j)
=
\{b_i\in\mathcal B
\mid
z_{i,j}^{\mathsf{int},(t)}=1\}$,
and the interim unmatched-BE set is
$\mathcal B_{\mathsf{unm}}^{\mathsf{int},(t)}
=
\{b_i\in\mathcal B
\mid
\sum_{s_j\in\mathcal S}
z_{i,j}^{\mathsf{int},(t)}=0\}$.
For each ES $s_j$, the aggregate scrubbing rate reserved through the
interim B\&E contracts is
$R_j^{S,\mathsf{book,int},(t)}
=
\sum_{b_i\in\mathcal B}
z_{i,j}^{\mathsf{int},(t)}
x_{i,j}^{B\leftrightarrow E,(t)}$.
The interim matching outcome, the associated B\&E main-service
contracts, aggregate booked rates, unmatched BEs, and the preserved
initial BE preference lists
$\{\mathcal L_i^{(t),0}\}_{b_i\in\mathcal B}$
are passed to the subsequent E\&C supplementary-reservation and final
risk-aware B\&E contract-adjustment phase.
The B\&E reserved rates, GNB-determined reservation prices, and
compensation terms remain fixed during the subsequent adjustment and
are not renegotiated.
The final contract configuration is determined only after the formal
E\&C supplementary reservations and aggregate shortfall risks have been
evaluated.

\subsubsection{E\&C Supplementary Reservation and Final Risk-Aware B\&E Contract Adjustment}
\label{subsubsec:es_cs_reservation_and_final_adjustment}

Following interim B\&E matching, ESs whose aggregate booked
scrubbing rates exceed their local physical capacities request
supplementary cloud capacity from the CS. Formal E\&C
reservation rates are allocated subject to the CS physical-capacity
constraint. The allocated rates are then used to
reevaluate each ES's aggregate shortfall risk and, when
necessary, perform risk-aware B\&E contract adjustment within
the affected submarket without global rematching.

\textbf{Step 1: Supplementary scrubbing-rate request calculation.}
Based on the interim B\&E matching outcome, each ES $s_j$ determines
its supplementary scrubbing-rate request as the excess of its aggregate
booked scrubbing rate over its local physical scrubbing-rate limit, i.e.,
$r_j^{S,\mathsf{req},(t)}
=
\big(
R_j^{S,\mathsf{book,int},(t)}-f_j^S
\big)^+
=
\big(
\sum_{b_i\in\mathcal B}
z_{i,j}^{\mathsf{int},(t)}
x_{i,j}^{B\leftrightarrow E,(t)}
-f_j^S
\big)^+$.
Here, $R_j^{S,\mathsf{book,int},(t)}$ is the aggregate reserved
scrubbing rate induced by the interim B\&E matching, whereas
$r_j^{S,\mathsf{req},(t)}$ represents the additional scrubbing rate
requested from the CS during the offline stage rather than realized
overflow traffic during online fulfillment. The aggregate supplementary
request is
$R^{S,\mathsf{req},(t)}
=
\sum_{s_j\in\mathcal S}r_j^{S,\mathsf{req},(t)}$.
Subject to its physical capacity constraint, the CS allocates
supplementary reserved rates
$\{x_j^{E\leftrightarrow C,(t)}\}_{s_j\in\mathcal S}$
as described next.

\textbf{Step 2: CS-side supplementary-capacity allocation and E\&C
contract generation.}
Based on the interim B\&E matching outcome, let
$\mathcal B_j^{\mathsf{int},(t)}
=
\mathcal{M}^{E\leftarrow B,\mathsf{int},(t)}(s_j)$
denote the set of BEs matched to ES $s_j$.
The interim reservation revenue received by $s_j$ is
$I_j^{S,\mathsf{int},(t)}
=
\sum_{b_i\in\mathcal B}
z_{i,j}^{\mathsf{int},(t)}
p_{i,j}^{B\leftrightarrow E,*,(t)}
x_{i,j}^{B\leftrightarrow E,(t)}
\Delta t$.
Let $\mathcal{G}_j^S\in(0,1]$ denote the fraction of this revenue that
$s_j$ is willing to spend on cloud-side supplementary reservation.
Its cloud-side reservation budget is
$B_j^{S,\mathsf{cld},(t)}
=
\mathcal{G}_j^S I_j^{S,\mathsf{int},(t)}$.
Assuming $c^{C,(t)}>0$, the maximum economically feasible supplementary
reserved rate is
$\overline r_j^{S,(t)}
=
\min\{
r_j^{S,\mathsf{req},(t)},
B_j^{S,\mathsf{cld},(t)}/(c^{C,(t)}\Delta t)
\}$.
If $\overline r_j^{S,(t)}=0$, ES $s_j$ has no effective cloud-side
reservation option other than zero allocation.
The feasible range $[0,\overline r_j^{S,(t)}]$ is discretized with
granularity $g_x>0$. The candidate supplementary reserved
rates are
$a_{j,k}
=
\min\{kg_x,\overline r_j^{S,(t)}\}$,
$k=0,1,\ldots,K_j$, where
$K_j=
\lceil\overline r_j^{S,(t)}/g_x\rceil$.
The zero level $a_{j,0}=0$ represents no cloud-side reservation.
For each positive candidate level, let
$A_{j,k}^{(t)}=a_{j,k}\Delta t$
denote the corresponding reserved scrubbing volume.
To quantify the economic value of supplementary cloud capacity to the
ES, define
$\mathcal R_{j,0}^{S,(t)}
=
\mathcal R_{j,\mathsf{sh}}^{S,(t)}
(\mathcal B_j^{\mathsf{int},(t)},0)$
and
$\mathcal R_{j,k}^{S,(t)}
=
\mathcal R_{j,\mathsf{sh}}^{S,(t)}
(\mathcal B_j^{\mathsf{int},(t)},a_{j,k})$.
The monetary-equivalent value of candidate level $a_{j,k}$ is
$V_{j,k}^{E\leftrightarrow C,(t)}
=
\mathcal{Y}_j^S
(\mathcal R_{j,0}^{S,(t)}-\mathcal R_{j,k}^{S,(t)})$.
For a unit reservation price $p$, the ES and CS surpluses are
$S_{j,k}^{S,E\leftrightarrow C,(t)}(p)
=
V_{j,k}^{E\leftrightarrow C,(t)}
-pA_{j,k}^{(t)}$
and
$S_{j,k}^{C,E\leftrightarrow C,(t)}(p)
=
(p-c^{C,(t)})A_{j,k}^{(t)}$,
respectively.
For each positive candidate level, the feasible unit-price interval is
$\underline p_{j,k}^{E\leftrightarrow C,(t)}=c^{C,(t)}$
and
$\overline p_{j,k}^{E\leftrightarrow C,(t)}
=
\min\{
V_{j,k}^{E\leftrightarrow C,(t)}/A_{j,k}^{(t)},
B_j^{S,\mathsf{cld},(t)}/A_{j,k}^{(t)}
\}$.
A positive candidate is price-feasible only if
$\underline p_{j,k}^{E\leftrightarrow C,(t)}
\leq
\overline p_{j,k}^{E\leftrightarrow C,(t)}$.
For each price-feasible candidate, the reservation price is determined
through generalized Nash bargaining:
\begin{equation}
\begin{aligned}
\max_{p}\quad&
\left[
S_{j,k}^{S,E\leftrightarrow C,(t)}(p)
\right]^{\mathcal{W}_j^{EC}}
\left[
S_{j,k}^{C,E\leftrightarrow C,(t)}(p)
\right]^{1-\mathcal{W}_j^{EC}}
\\
\mathrm{s.t.}\quad&
\underline p_{j,k}^{E\leftrightarrow C,(t)}
\leq p \leq
\overline p_{j,k}^{E\leftrightarrow C,(t)},
\end{aligned}
\label{eq:ec_nash_bargaining}
\end{equation}
where $\mathcal{W}_j^{EC}\in(0,1)$ denotes the bargaining weight of ES $s_j$.
The unconstrained Nash bargaining solution is
$p_{j,k}^{E\leftrightarrow C,\mathsf{NB},(t)}
=
(1-\mathcal{W}_j^{EC})
V_{j,k}^{E\leftrightarrow C,(t)}/A_{j,k}^{(t)}
+
\mathcal{W}_j^{EC}c^{C,(t)}$.
Accounting for the cloud-side reservation budget, the resulting
candidate unit price is
$
p_{j,k}^{E\leftrightarrow C,*,(t)}
=
\min
\left\{
p_{j,k}^{E\leftrightarrow C,\mathsf{NB},(t)},
\frac{B_j^{S,\mathsf{cld},(t)}}{A_{j,k}^{(t)}}
\right\}.
\label{eq:ec_contract_price}
$
For the zero-allocation level, set
$p_{j,0}^{E\leftrightarrow C,*,(t)}=0$.
Since realized cloud-side supplementary traffic is unavailable during
offline reservation, the CS evaluates each feasible candidate level
using the following proxy utility:
\begin{equation}
G_{j,k}^{C,(t)}
=
\left(
p_{j,k}^{E\leftrightarrow C,*,(t)}
-
c^{C,(t)}
\right)
a_{j,k}\Delta t.
\label{eq:cloud_proxy_utility_for_candidate_level}
\end{equation}
The discretized cloud-capacity occupancy is
$w_{j,k}=\lceil a_{j,k}/g_x\rceil$,
and the total discretized CS capacity is
$W^C=\lfloor f^C/g_x\rfloor$.

After indexing the ESs as
$s_1,\ldots,s_{|\mathcal S|}$, let $\mathcal D[m,w]$ denote the maximum
aggregate CS-side proxy utility obtained by selecting one candidate
level for each of the first $m$ ESs.
The discretized CS-side allocation problem has a
multiple-choice knapsack structure and is solved
using dynamic programming~\cite{pisinger1995minimal}.
With $\mathcal D[0,0]=0$ and
$\mathcal D[0,w]=-\infty$ for $w>0$, the state transition is
$
\mathcal D[m,w]
=
\max_{\substack{
0\leq k\leq K_m,\\
w_{m,k}\leq w
}}
\left\{
\mathcal D[m-1,w-w_{m,k}]
+
G_{m,k}^{C,(t)}
\right\}.
\label{eq:cloud_dp_transition}
$
After all state transitions,
$w^*\in
\arg\max_{0\leq w\leq W^C}
\mathcal D[|\mathcal S|,w]$
is selected, and the corresponding candidate index $k_j^*$ is recovered
for each ES by backtracking. If multiple solutions yield the same
aggregate proxy utility, the system first selects the one with the
larger total supplementary reserved rate and then breaks remaining ties
deterministically in ascending order of ES indices.
The formal supplementary reserved rate is
$x_j^{E\leftrightarrow C,(t)}=a_{j,k_j^*}$.
For $x_j^{E\leftrightarrow C,(t)}>0$, the corresponding unit reservation
price is
$p_j^{E\leftrightarrow C,(t)}
=
p_{j,k_j^*}^{E\leftrightarrow C,*,(t)}$,
and the formal E\&C supplementary contract is
$\mathbb C_j^{E\leftrightarrow C,(t)}
=
(x_j^{E\leftrightarrow C,(t)},
p_j^{E\leftrightarrow C,(t)})$.
Since $w_{j,k}$ is rounded upward, the DP allocation conservatively
satisfies the CS physical capacity constraint, and the discretization
gap decreases as $g_x$ decreases.

\textbf{Step 3: Final risk-aware B\&E contract adjustment.}
After the CS-side supplementary-capacity allocation is completed, the
formal E\&C supplementary reserved rate and unit reservation price,
$x_j^{E\leftrightarrow C,(t)}$ and
$p_j^{E\leftrightarrow C,(t)}$, are available for each ES $s_j$.
The system then reevaluates the aggregate shortfall risks of the interim
B\&E contract outcome. Initialize
$z_{i,j}^{\mathsf{cur},(t)}
=
z_{i,j}^{\mathsf{int},(t)}$
and
$\mathcal B_j^{\mathsf{cur},(t)}
=
\{b_i\in\mathcal B
\mid
z_{i,j}^{\mathsf{cur},(t)}=1\}$.
For each BE $b_i$, initialize the final rejection set as
$\mathcal S_i^{\mathsf{rej,fin},(t)}=\emptyset$.
The risk-excess ES set is defined as
$
\mathcal S_{\mathsf{risk}}^{(t)}
=
\left\{
s_j\in\mathcal S
\ \middle|\
\mathcal R_{j,\mathsf{sh}}^{S,(t)}
\left(
\mathcal B_j^{\mathsf{cur},(t)},
x_j^{E\leftrightarrow C,(t)}
\right)
>
\mathcal{Z}_j^S
\right\}.
\label{eq:risk_excess_es_set}
$
Contract adjustment is triggered only for ESs in
$\mathcal S_{\mathsf{risk}}^{(t)}$.
The ES-side proxy is updated by replacing the reference
E\&C rate--price pair with the formal pair
$(x_j^{E\leftrightarrow C,(t)},
p_j^{E\leftrightarrow C,(t)})$.
For compactness, define
\begin{equation}
\Pi_j^{S,\mathsf{upd},(t)}
(\mathcal B_j^{(t)})
=
\left.
\Pi_j^{S,\mathsf{int},(t)}
(\mathcal B_j^{(t)})
\right|_{\mathsf{ref}\to\mathsf{formal}}.
\label{eq:es_updated_proxy_score}
\end{equation}
For each
$s_j\in\mathcal S_{\mathsf{risk}}^{(t)}$
and
$b_i\in\mathcal B_j^{\mathsf{cur},(t)}$,
the marginal risk-adjusted retention score is
\begin{equation}
\begin{aligned}
\Phi_{i,j}^{S,\mathsf{rel},(t)}
={}&
\frac{
\Pi_j^{S,\mathsf{upd},(t)}
\left(
\mathcal B_j^{\mathsf{cur},(t)}
\right)
-
\Pi_j^{S,\mathsf{upd},(t)}
\left(
\mathcal B_j^{\mathsf{cur},(t)}
\setminus\{b_i\}
\right)
}{
x_{i,j}^{B\leftrightarrow E,(t)}
+
g_x
}
\\
&-
\mathcal{W}_j^S
\Bigg[
\mathcal R_{j,\mathsf{sh}}^{S,(t)}
\left(
\mathcal B_j^{\mathsf{cur},(t)},
x_j^{E\leftrightarrow C,(t)}
\right)
\\[-1mm]
&\hspace{3.2em}
-
\mathcal R_{j,\mathsf{sh}}^{S,(t)}
\left(
\mathcal B_j^{\mathsf{cur},(t)}
\setminus\{b_i\},
x_j^{E\leftrightarrow C,(t)}
\right)
\Bigg]^+ .
\end{aligned}
\label{eq:marginal_risk_adjusted_retention_score}
\end{equation}
Here, $\mathcal{W}_j^S\geq0$ controls the penalty assigned to the marginal
shortfall-risk contribution of a retained contract. A smaller
$\Phi_{i,j}^{S,\mathsf{rel},(t)}$ indicates a lower marginal retention
value per unit of reserved scrubbing rate, a larger contribution to the
aggregate shortfall risk, or both.
For each risk-excess ES, the system releases contracts in ascending
order of $\Phi_{i,j}^{S,\mathsf{rel},(t)}$ until its retained BE set
satisfies the aggregate shortfall-risk threshold. Each released BE is
added to the affected-BE set, while the releasing ES is recorded in that
BE's final rejection set for the remainder of the current adjustment.
Let $\mathcal B_{\mathsf{aff}}^{(t)}$ denote the affected-BE set,
consisting of the released BEs and those unmatched after the interim
matching.
Given the formal E\&C supplementary reserved rate
$x_j^{E\leftrightarrow C,(t)}$, let
$\mathfrak F_j^{S,\mathsf{fin},(t)}$
denote the family of final risk-feasible acceptance sets of ES $s_j$,
i.e., those satisfying the concurrency, controlled-overbooking, and
aggregate shortfall-risk constraints. The empty set is included in
$\mathfrak F_j^{S,\mathsf{fin},(t)}$ and represents the no-contract
outcome, with
$\Pi_j^{S,\mathsf{upd},(t)}(\emptyset)=0$.
For each affected BE $b_i$, the final reapplication list is
\(
\mathcal L_i^{\mathsf{fin},(t)}
=
\Big\{
s_j \in
\mathcal L_i^{(t),0}
\setminus
\mathcal S_i^{\mathsf{rej,fin},(t)}
:
\mathcal B_j^{\mathsf{cur},(t)}
\cup\{b_i\}
\in
\mathfrak F_j^{S,\mathsf{fin},(t)}
\Big\}
\)
Each affected BE may then reapply to the ESs in its final admissible
list according to its preserved initial preference order.
For each ES $s_j$ receiving final applications, let
$\mathcal B_j^{\mathsf{app,fin},(t)}$
denote the newly applying BEs and define
$\mathcal B_j^{\mathsf{cand,fin},(t)}
=
\mathcal B_j^{\mathsf{cur},(t)}
\cup
\mathcal B_j^{\mathsf{app,fin},(t)}$.
Given the formal E\&C reservation, ES $s_j$ selects a final
risk-feasible acceptance set according to
$\mathcal B_j^{\mathsf{fin},(t),*}
\in
\arg\max_{\mathcal B_j^{(t)}
\subseteq
\mathcal B_j^{\mathsf{cand,fin},(t)},
\,
\mathcal B_j^{(t)}
\in
\mathfrak F_j^{S,\mathsf{fin},(t)}}
\Pi_j^{S,\mathsf{upd},(t)}
\left(
\mathcal B_j^{(t)}
\right)$.
The corresponding choice function is
$\operatorname{Ch}_j^{S,\mathsf{fin},(t)}
(\mathcal B_j^{\mathsf{cand,fin},(t)})
=
\mathcal B_j^{\mathsf{fin},(t),*}$, evaluated exactly over the
discretized candidate sets.
If multiple acceptance sets attain the same updated proxy score,
$s_j$ first prefers the set with the larger aggregate B\&E reservation
revenue and then applies a fixed lexicographic rule over BE indices.
The final reapplication follows the same application--rejection dynamics
as the interim B\&E matching process. Each rejected BE removes the
rejecting ES from its final admissible list, records it in
$\mathcal S_i^{\mathsf{rej,fin},(t)}$, and, if the list remains
nonempty, applies to its next most-preferred ES. The set
$\mathcal S_i^{\mathsf{rej,fin},(t)}$ is preserved across outer
payment-feasibility iterations. The procedure terminates when no
affected BE remains unmatched with a nonempty final admissible list.
Upon termination, every retained ES satisfies the
concurrency, controlled-overbooking, and aggregate shortfall-risk
constraints.

\textbf{Step 4: Payment-feasibility check and final output.}
After the final risk-aware B\&E contract adjustment, the system
rechecks the payment feasibility of each retained E\&C supplementary
contract. This check is necessary because the B\&E adjustment may reduce
the reservation revenue of an ES and thereby affect its ability to
afford the previously reserved cloud-side capacity.
For each ES $s_j$ with
$x_j^{E\leftrightarrow C,(t)}>0$,
the system recomputes its current B\&E reservation revenue
$I_j^{S,\mathsf{fin},(t)}$ from the retained B\&E contracts and verifies
the payment-feasibility condition
$p_j^{E\leftrightarrow C,(t)}x_j^{E\leftrightarrow C,(t)}\Delta t
\leq\mathcal{G}_j^S I_j^{S,\mathsf{fin},(t)}$.
If this condition fails, the E\&C contract is canceled by setting
$x_j^{E\leftrightarrow C,(t)}=0$ and $p_j^{E\leftrightarrow C,(t)}=0$.
The system then recomputes the aggregate shortfall risks under the
updated supplementary-reservation outcome, updates
$\mathcal S_{\mathsf{risk}}^{(t)}$, and repeats the risk-aware B\&E
contract adjustment from the current B\&E outcome without resetting it
to the interim matching.
The payment-feasibility check and
risk-aware adjustment are repeated until all retained E\&C contracts
are payment-feasible.
Finally, set
$z_{i,j}^{(t)}=z_{i,j}^{\mathsf{cur},(t)}$
for all $(b_i,s_j)\in\mathcal B\times\mathcal S$.
The final unmatched-BE set is
$\mathcal B_{\mathsf{unm}}^{(t)}
=
\{b_i\in\mathcal B
\mid
\sum_{s_j\in\mathcal S}z_{i,j}^{(t)}=0\}$.
The mechanism outputs the final B\&E matching indicators
$\{z_{i,j}^{(t)}\}$, the retained B\&E main-service contracts,
the retained formal E\&C supplementary contracts, and
$\mathcal B_{\mathsf{unm}}^{(t)}$.
The final B\&E contracts are not required to be deterministically
covered by the combined local and cloud-side scrubbing rates.
Instead, they are retained subject to the maximum controlled-overbooking
and aggregate shortfall-risk constraints, thereby balancing resource
utilization and fulfillment risk.

\begin{algorithm}[!t]
\caption{E\&C Supplementary Reservation and Final Risk-Aware B\&E Contract Adjustment}
\label{alg:escs_rematching}
\begin{algorithmic}[1]

\STATE \textbf{Input:}
interim B\&E outcome $\{z_{i,j}^{\mathsf{int},(t)}\}$,
booked rates $\{R_j^{S,\mathsf{book,int},(t)}\}$,
and initial preference lists $\{\mathcal L_i^{(t),0}\}$.

\STATE Compute each ES request $r_j^{S,\mathsf{req},(t)}$,
cloud budget $B_j^{S,\mathsf{cld},(t)}$, and feasible upper rate
$\overline r_j^{S,(t)}$.

\STATE Generate candidate E\&C reservation levels $\{a_{j,k}\}$
and determine GNB prices
$\{p_{j,k}^{E\leftrightarrow C,*,(t)}\}$.

\STATE Compute CS-side proxy utilities $\{G_{j,k}^{C,(t)}\}$
and solve the multiple-choice knapsack problem via DP.

\STATE Establish formal E\&C contracts with
$x_j^{E\leftrightarrow C,(t)}\leftarrow a_{j,k_j^*}$.

\STATE Initialize
$z_{i,j}^{\mathsf{cur},(t)}
\leftarrow
z_{i,j}^{\mathsf{int},(t)}$,
$\mathcal S_i^{\mathsf{rej,fin},(t)}\leftarrow\emptyset$,
and the affected-BE set with interim-unmatched BEs.

\REPEAT

\STATE Compute the risk-excess ES set
$\mathcal S_{\mathsf{risk}}^{(t)}$.

\WHILE{$\mathcal S_{\mathsf{risk}}^{(t)}\neq\emptyset$}

\STATE Release B\&E contracts in ascending order of
$\Phi_{i,j}^{S,\mathsf{rel},(t)}$ until risk-feasible;
add released BEs to the affected set.

\STATE Update $\mathcal S_{\mathsf{risk}}^{(t)}$.

\ENDWHILE

\STATE Perform localized reapplication for affected BEs
using $\mathcal L_i^{\mathsf{fin},(t)}$; record every final-stage
rejection in $\mathcal S_i^{\mathsf{rej,fin},(t)}$ and never reset it;
update the current B\&E outcome.

\STATE Recheck E\&C payment feasibility; cancel infeasible
E\&C contracts and recompute shortfall risks.

\UNTIL{no E\&C contract is canceled in the current iteration}

\STATE \textbf{Output:}
final B\&E indicators $\{z_{i,j}^{(t)}\}$,
retained B\&E and E\&C contracts, and
$\mathcal B_{\mathsf{unm}}^{(t)}$.

\end{algorithmic}
\end{algorithm}

\subsection{Online Contract Fulfillment}
\label{subsec:online_contract_fulfillment}

After demand realization, each ES fulfills finalized B\&E
contracts using its local scrubbing capacity first, followed
by its formally reserved E\&C capacity for remaining
contractual demand. WPA is applied at either layer when aggregate contractual demand
at that layer exceeds its available capacity. No contract
rematching, price renegotiation, or additional cloud
reservation is permitted during online fulfillment.

\textbf{Step 1: Contractual fulfillment demand determination.}
For each ES $s_j$, let
$\mathcal F_j^{(t)}
=
\{b_i\in\mathcal B\mid z_{i,j}^{(t)}=1\}$
denote the set of BEs associated with effective B\&E contracts.
For each $b_i\in\mathcal F_j^{(t)}$, the contractual fulfillment demand
is
$h_{i,j}^{(t)}
=
\min\{
d_i^{B,(t)},
x_{i,j}^{B\leftrightarrow E,(t)}\Delta t
\}$.
Thus, $h_{i,j}^{(t)}$ represents the portion of realized suspicious
traffic falling within the reserved volume of the corresponding B\&E
contract. Traffic exceeding the reserved volume lies outside the
contractual fulfillment demand and does not trigger compensation.

\textbf{Step 2: Weighted proportional allocation.}
When multiple contracted BEs compete for limited scrubbing capacity, the
system assigns BE $b_i$ the fulfillment weight
$\mathcal{O}_{i,j}^{(t)}
=
q_{i,j}^{E\rightarrow B,(t)}
+
\mathcal{X} p_{i,j}^{B\leftrightarrow E,*,(t)}$,
where $\mathcal{X}\geq0$ controls the relative emphasis on compensation
liability and reservation revenue. A larger $\mathcal{O}_{i,j}^{(t)}$
indicates a higher fulfillment priority.
Given demands $\{u_i\}$, available capacity $C$, and weights
$\{\mathcal{O}_i\}$, the WPA operator allocates capacity proportionally to
the weighted demands $\{\mathcal{O}_i u_i\}$ when the aggregate demand
exceeds $C$. Any allocation exceeding its demand is truncated, and the
remaining capacity is redistributed among unsatisfied BEs using the
same rule. If the aggregate weighted demand of the remaining BEs is
zero, the remaining capacity is allocated proportionally to their
unweighted demands. The procedure continues until all demands are
satisfied or the available capacity is exhausted. We denote the
resulting allocation by
$
\{a_i\}
=
\operatorname{WPA}
\left(
\{u_i\},
C,
\{\mathcal{O}_i\}
\right).
$
The WPA allocation satisfies
$0\leq a_i\leq u_i$ and
$\sum_i a_i=\min\{C,\sum_i u_i\}$.
The same operator is used for both edge-side and cloud-side fulfillment.

\textbf{Step 3: Edge-side local fulfillment.}
For each ES $s_j$, its local physical scrubbing capacity
$f_j^S\Delta t$ is first allocated among
$\mathcal F_j^{(t)}$ according to
\(
\{a_{i,j}^{S,(t)}\}_{b_i\in\mathcal F_j^{(t)}}
=
\operatorname{WPA}\bigl(
\{h_{i,j}^{(t)}\}_{b_i\in\mathcal F_j^{(t)}},
\allowbreak
f_j^S\Delta t,
\allowbreak
\{\mathcal{O}_{i,j}^{(t)}\}_{b_i\in\mathcal F_j^{(t)}}
\bigr)
\).
The realized edge-side scrubbing volume is
\(y_{i,j}^{B\leftrightarrow E,(t)}
=a_{i,j}^{S,(t)}\).
and the remaining contractual demand after edge fulfillment is
$d_{i,j}^{\mathsf{rem},(t)}
=
h_{i,j}^{(t)}
-
y_{i,j}^{B\leftrightarrow E,(t)}$.

\textbf{Step 4: Cloud-side supplementary fulfillment.}
The remaining contractual demand is subsequently served using the
formal E\&C supplementary reservation of ES $s_j$:
\(
\{a_{i,j}^{C,(t)}\}_{b_i\in\mathcal F_j^{(t)}}
=
\operatorname{WPA}\bigl(
\{d_{i,j}^{\mathsf{rem},(t)}\}_{b_i\in\mathcal F_j^{(t)}},
\allowbreak
x_j^{E\leftrightarrow C,(t)}\Delta t,
\allowbreak
\{\mathcal{O}_{i,j}^{(t)}\}_{b_i\in\mathcal F_j^{(t)}}
\bigr)
\).
The realized cloud-side supplementary scrubbing volume is
$y_{i,j}^{E\leftrightarrow C,(t)}
=a_{i,j}^{C,(t)}$.
If no effective E\&C supplementary contract is retained for $s_j$, then
$x_j^{E\leftrightarrow C,(t)}=0$ and hence
$y_{i,j}^{E\leftrightarrow C,(t)}=0$
for all $b_i\in\mathcal F_j^{(t)}$.
The resulting total scrubbing volume, uncovered suspicious
traffic, and scrubbing coverage ratio follow the definitions
in Section~\ref{subsubsec:be_utility_risk}.

\section{Performance Evaluation}
\label{sec:performance_evaluation}

\subsection{Experimental Setup}
\label{subsec:experimental_setup}

\subsubsection{Dataset and Prediction Settings}
\label{subsubsec:dataset_settings}
We derive traffic-demand proxies from CICDDoS2019~\cite{sharafaldin2019developing} by aggregating bidirectional packet-length fields into 5-s timeslots by destination IP, port, and protocol. The pool contains benign-only and attack-containing traces; because no attack-label filtering is applied, these volumes do not represent attack-only traffic. Three traces are cyclically assigned to BEs and transformed by scaling, stable-regime alignment, and temporal perturbations. Each trace is split into 3,452 training, 1,150 validation, and 1,152 test timeslots; evaluation uses 20 consecutive test slots.

Our lightweight Transformer uses a 12-slot context, one encoder layer, two attention heads, and a 16-dimensional hidden representation, and is trained once per BE per evaluation horizon. Forecasting and offline contracting at slot $t$ use observations through $t-1$; realized demand enters the history only after fulfillment. Risk estimation uses 64 validation-residual scenarios fixed over the horizon.

\subsubsection{Benchmark Methods}
We compare \textbf{Proposed} with four benchmarks: \textbf{Greedy}, 
\textbf{On2TM}, \textbf{PAO-adapted}, and \textbf{Qi-hybrid}. 
Their demand information, decision timing, and constraints are summarized 
in Table~\ref{tab:comparison_methods}. Since these methods differ in 
information availability and decision structure, the comparisons 
characterize performance under distinct operating models rather than 
identical optimization problems.

Five ablations isolate the components of Proposed: \textbf{NoTrans} uses a 12-slot moving average; \textbf{LinearPrice} replaces GNB pricing in both contractual layers with linear interpolation; \textbf{NoOB} limits ES reservations to physical capacity; \textbf{NoCloud} removes E\&C supplementary reservation; and \textbf{NoRisk} omits final risk-aware B\&E adjustment while retaining earlier contract-formation procedures.

\begin{table*}[!t]
\centering
\caption{Comparison methods and decision characteristics.}
\label{tab:comparison_methods}
\footnotesize
\setlength{\tabcolsep}{4pt}
\renewcommand{\arraystretch}{1.10}
\begin{tabular}{@{}p{0.15\textwidth}p{0.17\textwidth}p{0.26\textwidth}p{0.33\textwidth}@{}}
\hline
\textbf{Method} & \textbf{Demand Information} &
\textbf{Decision Structure} &
\textbf{Constraints and Scope} \\
\hline
Proposed & Causal one-step forecasts &
Ex-ante B\&E/E\&C contracts; online WPA &
Budgets, pricing, controlled rate overbooking,
aggregate ES risk, and cloud capacity \\
Greedy & Realized demand &
Direct online edge/cloud allocation &
Physical capacity and connectivity;
no ex-ante contracts or BE budget enforcement \\
On2TM & Realized demand &
Online B\&E matching and E\&C reservation &
Pricing, budgets, capacity, and interim
overbooking; no final risk adjustment \\
PAO-adapted & Causal forecasts and realized demand &
Five-slot planning and online allocation &
Capacity and allocation costs;
no B\&E/E\&C contractual risk constraints \\
Qi-hybrid & Ex-ante forecasts and realized demand
at spot stage &
Adapted futures contracting and online
spot matching with atomic tasks &
Adapted OA-style overbooking and risk constraints;
deterministic contractual attendance \\
\hline
\end{tabular}
\end{table*}

\subsubsection{Evaluation Metrics and Parameters}
\label{subsubsec:evaluation_metrics}

We evaluate the scrubbing coverage ratio (SCR) and residual traffic 
ratio (RTR) for processed and uncovered demand, respectively, and 
edge/cloud resource utilization (ERU/CRU) for physical capacity usage. 
For contractual variants, we additionally evaluate the contract 
admission ratio (CAR), realized social welfare (SW) summed over BEs, 
ESs, and CS, and the maximum estimated aggregate ES shortfall risk. 
Computational efficiency is measured by wall-clock time under the 
corresponding timing boundaries.

Table~\ref{tab:simulation_parameters} summarizes the simulation
settings. For each BE scale, the number of ESs is set to
$|\mathcal S|=\max\{3,\min\{6,\lceil|\mathcal B|/8\rceil\}\}$.
Each of the 20 random seeds covers 20 consecutive 5-s slots,
with compared methods sharing the same trace-derived demand
and physical market instance for each seed. Horizon-level
metrics are summarized across seeds with 95\% confidence
intervals.

\begin{table}[!t]
\centering
\caption{Simulation parameters.}
\label{tab:simulation_parameters}
\footnotesize
\setlength{\tabcolsep}{4pt}
\renewcommand{\arraystretch}{1.08}
\begin{tabular}{@{}p{0.55\columnwidth}p{0.35\columnwidth}@{}}
\hline
\textbf{Parameter} & \textbf{Value} \\
\hline
Number of BEs, $|\mathcal B|$ & $\{20,30,40,50,60\}$ \\
Number of ESs, $|\mathcal S|$ & $\{3,4,5,6,6\}$ \\
Number of CSs & $1$ \\
Evaluation horizon & $20$ timeslots \\
Timeslot duration, $\Delta t$ & $5$ s \\
Random seeds & $20$ \\
Nominal target traffic P95 & $[2,6]$ Gbit \\
Scale jitter & $[0.85,1.15]$ \\
Per-slot perturbation & $[0.97,1.03]$ \\
Maximum ES overbooking ratio & $1.35$ \\
ES shortfall-risk threshold & $0.1$ \\
Empirical residual scenarios & $64$ \\
Cloud budget fraction, $\mathcal{G}$ & $0.03$ \\
B\&E/E\&C GNB weights & $0.5/0.5$ \\
Maximum E\&C candidate levels & $25$ (including zero) \\
\hline
\end{tabular}
\end{table}

\subsection{Overall Performance}
\label{subsec:overall_performance}

\begin{figure}[!t]
    \centering
    \includegraphics[width=0.75\columnwidth]{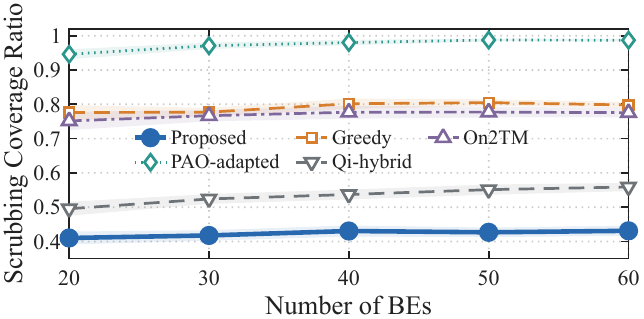}
    \caption{Overall scrubbing coverage ratio across different market scales.}
    \label{fig:overall_scr}
\end{figure}

\begin{figure}[!t]
    \centering
    \includegraphics[width=\columnwidth]{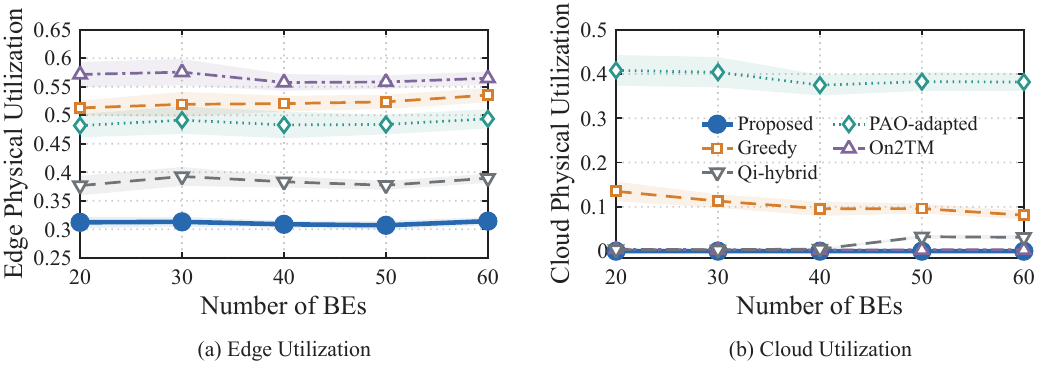}
    \caption{Physical resource utilization: (a) edge utilization
    and (b) cloud utilization.}
    \label{fig:resource_utilization}
\end{figure}

Fig.~\ref{fig:overall_scr} shows that Proposed achieves an
SCR of approximately $0.41$--$0.43$, compared with
$0.78$--$0.81$ for Greedy and $0.50$--$0.56$ for
Qi-hybrid. Unlike Proposed, which fulfills pre-established
contracts within their reserved volumes, Greedy allocates
resources using realized demand, while Qi-hybrid permits
realized-demand spot matching
(Table~\ref{tab:comparison_methods}). These comparisons
reflect different information and contractual decision
structures rather than identical optimization problems.

Fig.~\ref{fig:resource_utilization} reports physical ERU
and CRU. Proposed processes nearly all fulfilled traffic
at the edge, with near-zero realized cloud utilization.
Under edge-first fulfillment, cloud-side processing is
limited by residual contractual demand and finalized
E\&C reserved capacity. These results indicate limited
realized cloud participation under the evaluated settings,
consistent with the edge-first allocation policy and
ex-ante reservation constraints.

\subsection{Ablation Studies}
\label{subsec:ablation_studies}

\begin{figure}[!t]
    \centering
    \includegraphics[width=\columnwidth]{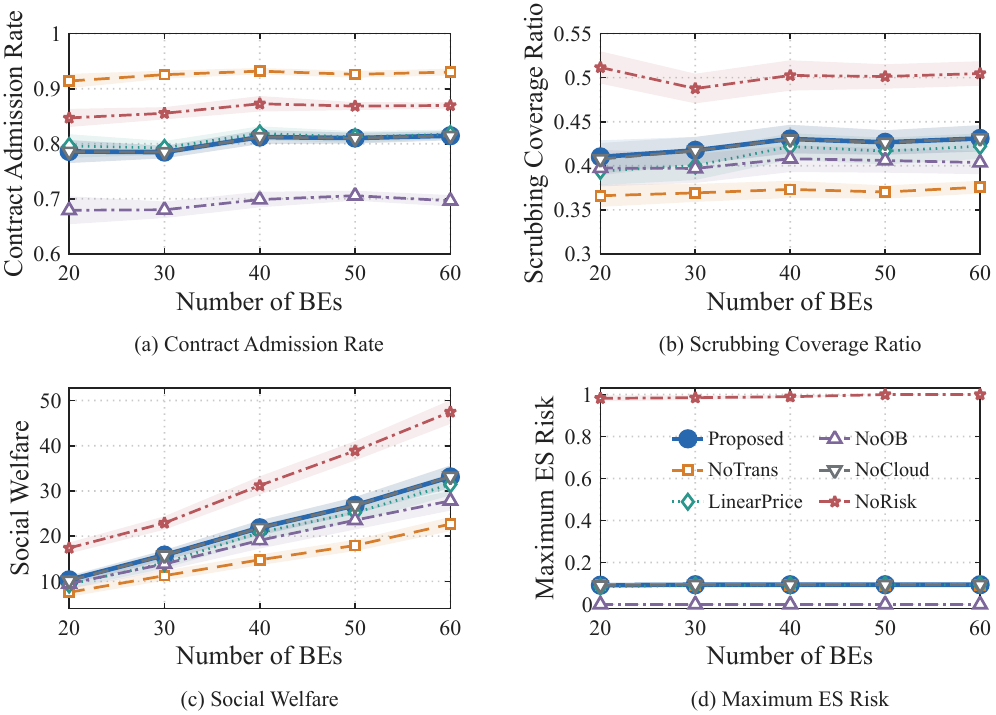}
    \caption{Ablation study: (a) contract admission rate,
    (b) scrubbing coverage ratio, (c) social welfare, and
    (d) maximum empirical aggregate ES shortfall risk. All
    variants except NoRisk satisfy the prescribed ex-ante
    empirical risk constraint under the evaluated settings.}
    \label{fig:ablation_studies}
\end{figure}

\begin{figure}[!t]
    \centering
    \includegraphics[width=\columnwidth]{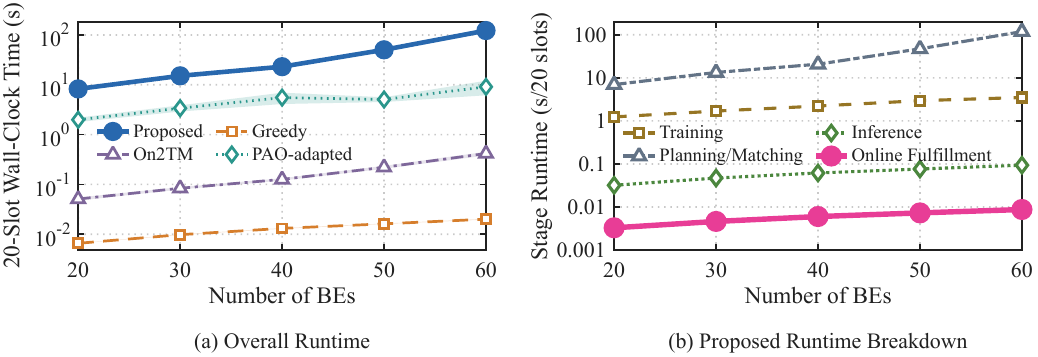}
    \caption{Computational runtime: (a) method-specific
    wall-clock time over a 20-slot horizon and (b) mean
    runtime decomposition of the proposed mechanism.
    Offline planning and matching dominate the proposed
    mechanism's computational cost, whereas online
    fulfillment incurs substantially lower cumulative
    overhead.}
    \label{fig:computational_efficiency}
\end{figure}

Fig.~\ref{fig:ablation_studies}(a)--(b) distinguishes contract admission from realized coverage. NoTrans admits more BEs than Proposed but attains lower SCR in the plotted results, demonstrating that admission counts alone do not determine fulfilled traffic volume. The effects of NoOB and LinearPrice further reflect the roles of controlled overbooking and contractual pricing in selecting feasible service commitments. Proposed and NoCloud have similar SCR, with mean differences of approximately $-0.0003$ to $0.0026$ across scales. Although little cloud traffic is processed, supplementary reservations may still alter final B\&E matching; the NoCloud comparison is therefore not solely a measure of cloud-side traffic volume.

Fig.~\ref{fig:ablation_studies}(c) compares realized SW per slot under common utility definitions. Pricing affects welfare indirectly through contract feasibility and service allocation, even though internal transfers cancel in aggregate welfare. NoRisk attains the highest SW in the plotted settings, but its apparent welfare advantage must be interpreted alongside its violation of the prescribed empirical risk constraint.

Fig.~\ref{fig:ablation_studies}(d) shows the clearest risk tradeoff. With the empirical shortfall-risk threshold set to $\mathcal{Z}=0.1$, Proposed and all ablations except NoRisk pass the recorded risk-feasibility checks for all 100 evaluated horizon-instance pairs per method, whereas NoRisk fails all 100. NoRisk disables only the \emph{final} risk-aware adjustment; earlier risk-related procedures remain. These are violations of an ex-ante empirical shortfall-risk constraint, not observed service failures or a distribution-free probability guarantee.

\subsection{Computational Efficiency}
\label{subsec:computational_efficiency}
Fig.~\ref{fig:computational_efficiency}(a) compares the four methods with a common overall timing boundary over 20 slots, after shared physical-instance construction. Proposed incurs substantial computation from predictor training and repeated contract planning. Qi Hybrid is excluded from this wall-clock comparison because its adapted futures/spot timing scope is not directly harmonized with the four canonical implementations.

Fig.~\ref{fig:computational_efficiency}(b) decomposes
Proposed runtime into training, inference, offline
planning/matching, and online fulfillment. Offline
planning/matching dominates the measured runtime and
increases sharply with market size. By contrast, online
WPA operates on fixed contracts without rematching or
renegotiation, incurring substantially lower cumulative
runtime. However, since contracts are reconstructed for
each target slot, this computational separation does not
establish an end-to-end real-time guarantee.

\section{Conclusion}
\label{sec:conclusion}

This paper studied ex-ante service provisioning in
cloud--edge DDoS scrubbing markets under uncertain demand.
We developed a prediction-driven two-stage mechanism
combining interim B\&E matching with controlled
overbooking, E\&C supplementary reservation, final
risk-aware B\&E adjustment, and online fulfillment.
Finalized E\&C reservations inform the final B\&E
adjustment, enabling cross-layer contract coordination.
Trace-derived experiments showed that Proposed satisfied
the prescribed empirical shortfall-risk constraint,
whereas NoRisk violated it under the evaluated settings.
Coverage was primarily limited by contract admission
and reserved volumes rather than contractual fulfillment
shortfalls. Cloud utilization remained limited, while
offline planning dominated runtime despite low online
fulfillment overhead. Future work will address contract
admission, cloud participation, and scalable offline planning.

\bibliographystyle{IEEEtran}
\bibliography{references}

@article{zargar2013survey,
  author  = {Saman Taghavi Zargar and James Joshi and David Tipper},
  title   = {A Survey of Defense Mechanisms Against Distributed Denial of Service ({DDoS}) Flooding Attacks},
  journal = {IEEE Communications Surveys \& Tutorials},
  year    = {2013},
  volume  = {15},
  number  = {4},
  pages   = {2046--2069},
  doi     = {10.1109/SURV.2013.031413.00127}
}

@article{yan2016sdn,
  author  = {Qiao Yan and F. Richard Yu and Qingxiang Gong and Jianqiang Li},
  title   = {Software-Defined Networking ({SDN}) and Distributed Denial of Service ({DDoS}) Attacks in Cloud Computing Environments: A Survey, Some Research Issues, and Challenges},
  journal = {IEEE Communications Surveys \& Tutorials},
  year    = {2016},
  volume  = {18},
  number  = {1},
  pages   = {602--622},
  doi     = {10.1109/COMST.2015.2487361}
}

@article{li2021optimal,
  author  = {Teng Li},
  title   = {Optimal Cloud Assistance Policy of End-Edge-Cloud Ecosystem for Mitigating Edge Distributed Denial of Service Attacks},
  journal = {Journal of Cloud Computing},
  year    = {2021},
  volume  = {10},
  number  = {1},
  articleno = {42},
  doi     = {10.1186/s13677-021-00257-3}
}

@article{chen2020traffic,
  author  = {Min Chen and Yiming Miao and Hamid Gharavi
             and Long Hu and Iztok Humar},
  title   = {Intelligent Traffic Adaptive Resource Allocation
             for Edge Computing-Based 5G Networks},
  journal = {IEEE Transactions on Cognitive Communications
             and Networking},
  volume  = {6},
  number  = {2},
  pages   = {499--508},
  year    = {2020},
  doi     = {10.1109/TCCN.2019.2953061}
}

@article{graph2025survey,
  author  = {Yanpeng Dai and Ling Lyu and Nan Cheng
             and Min Sheng and Junyu Liu and Xiucheng Wang
             and Shuguang Cui and Lin Cai and Xuemin Shen},
  title   = {A Survey of Graph-Based Resource Management
             in Wireless Networks---Part I:
             Optimization Approaches},
  journal = {IEEE Transactions on Cognitive Communications
             and Networking},
  volume  = {11},
  number  = {4},
  pages   = {2078--2100},
  year    = {2025},
  doi     = {10.1109/TCCN.2024.3508783}
}

@article{transformer2025cloudedge,
  author  = {Shijing Yuan and Xiang Chen and Suchuan Xing
             and Jie Li and Hongyang Chen and Zhi Liu
             and Song Guo},
  title   = {Transformer-Based Scalable Multi-Agent
             Reinforcement Learning for Joint Resource
             Optimization in Cloud--Edge--End Video
             Streaming Systems},
  journal = {IEEE Transactions on Cognitive Communications
             and Networking},
  volume  = {12},
  pages   = {3482--3496},
  year    = {2026},
  doi     = {10.1109/TCCN.2025.3612689}
}

@inproceedings{jiao2019online,
  author    = {Lei Jiao and Ruiting Zhou and Xiaojun Lin and Xu Chen},
  title     = {Online Scheduling of Traffic Diversion and Cloud Scrubbing with Uncertainty in Current Inputs},
  booktitle = {Proceedings of the Twentieth ACM International Symposium on Mobile Ad Hoc Networking and Computing},
  year      = {2019},
  pages     = {271--280},
  doi       = {10.1145/3323679.3326525}
}

@inproceedings{you2020scheduling,
  author    = {Wencong You and Lei Jiao and Jun Li and Ruiting Zhou},
  title     = {Scheduling {DDoS} Cloud Scrubbing in {ISP} Networks via Randomized Online Auctions},
  booktitle = {IEEE INFOCOM 2020 -- IEEE Conference on Computer Communications},
  year      = {2020},
  pages     = {1658--1667},
  doi       = {10.1109/INFOCOM41043.2020.9155493}
}

@article{he2022game,
  author  = {Qiang He and Cheng Wang and Guangming Cui and Bo Li and Rui Zhou and Qingguo Zhou and Yang Xiang and Hai Jin and Yun Yang},
  title   = {A Game-Theoretical Approach for Mitigating Edge {DDoS} Attack},
  journal = {IEEE Transactions on Dependable and Secure Computing},
  year    = {2022},
  volume  = {19},
  number  = {4},
  pages   = {2333--2348},
  doi     = {10.1109/TDSC.2021.3055559}
}

@article{zhou2024online,
  author  = {Ruiting Zhou and Yifan Zeng and Lei Jiao and Yi Zhong and Liujing Song},
  title   = {Online and Predictive Coordinated Cloud-Edge Scrubbing for {DDoS} Mitigation},
  journal = {IEEE Transactions on Mobile Computing},
  year    = {2024},
  volume  = {23},
  number  = {10},
  pages   = {9208--9223},
  doi     = {10.1109/TMC.2024.3360077}
}

@article{liwang2022unifying,
  author  = {Minghui Liwang and Ruitao Chen and Xianbin Wang and Xuemin Shen},
  title   = {Unifying Futures and Spot Market: Overbooking-Enabled Resource Trading in Mobile Edge Networks},
  journal = {IEEE Transactions on Wireless Communications},
  volume  = {21},
  number  = {7},
  pages   = {5467--5485},
  year    = {2022},
  doi     = {10.1109/TWC.2022.3141094}
}

@article{deng2022resource,
  author  = {Yuchuan Deng and Hao Jiang and Peijing Cai and Tong Wu and
             Pan Zhou and Beibei Li and Hao Lu and Jing Wu and Xin Chen and
             Kehao Wang},
  title   = {Resource Provisioning for Mitigating Edge {DDoS} Attacks
             in {MEC}-Enabled {SDVN}},
  journal = {IEEE Internet of Things Journal},
  volume  = {9},
  number  = {23},
  pages   = {24264--24280},
  year    = {2022},
  doi     = {10.1109/JIOT.2022.3189975}
}

@article{wu2026cooperative,
  author  = {Yifei Wu and Fan Yang and Qian Ma and Guocheng Liao and Xu Chen},
  title   = {Distributed Cooperative Defense Against {DDoS} Attacks in
             Edge-Cloud Computing Networks: A Game-Theoretic Approach},
  journal = {IEEE Transactions on Networking},
  volume  = {34},
  pages   = {3164--3178},
  year    = {2026},
  doi     = {10.1109/TON.2026.3659868}
}

@article{neira2023intelligent,
  author  = {Anderson Bergamini de Neira and
             Alex Medeiros de Ara{\'u}jo and
             Michele Nogueira},
  title   = {An Intelligent System for {DDoS} Attack Prediction
             Based on Early Warning Signals},
  journal = {IEEE Transactions on Network and Service Management},
  volume  = {20},
  number  = {2},
  pages   = {1254--1266},
  year    = {2023},
  doi     = {10.1109/TNSM.2022.3223881}
}

@article{qiu2022auction,
  author  = {Houming Qiu and Kun Zhu and Nguyen Cong Luong
             and Changyan Yi and Dusit Niyato and Dong In Kim},
  title   = {Applications of Auction and Mechanism Design
             in Edge Computing: A Survey},
  journal = {IEEE Transactions on Cognitive Communications
             and Networking},
  volume  = {8},
  number  = {2},
  pages   = {1034--1058},
  year    = {2022},
  doi     = {10.1109/TCCN.2022.3147196}
}

@article{pan2025auction,
  author  = {Jie Pan and Jintao Wu and Yiwen Zhang},
  title   = {An Online Game-Integrated Auction Scheme for {DDoS} Mitigation in Mobile Edge Computing},
  journal = {IEEE Transactions on Dependable and Secure Computing},
  volume  = {22},
  number  = {6},
  pages   = {7665--7679},
  year    = {2025},
  doi     = {10.1109/TDSC.2025.3599552}
}

@article{lee2024ips,
  author  = {Chun-Yen Lee and Chia-Hung Lin and Zhan-Lun Chang and
             Chih-Yu Wang and Hung-Yu Wei},
  title   = {Joint Resource Allocation and Intrusion Prevention
             System Deployment for Edge Computing},
  journal = {IEEE Transactions on Services Computing},
  volume  = {17},
  number  = {5},
  pages   = {2502--2515},
  year    = {2024},
  doi     = {10.1109/TSC.2024.3441313}
}

@article{you2024incentive,
  author  = {Feiran You and Xin Yuan and Wei Ni
             and Abbas Jamalipour},
  title   = {A Novel Privacy-Preserving Incentive Mechanism
             for Multi-Access Edge Computing},
  journal = {IEEE Transactions on Cognitive Communications
             and Networking},
  volume  = {10},
  number  = {5},
  pages   = {1928--1943},
  year    = {2024},
  doi     = {10.1109/TCCN.2024.3391303}
}

@article{wang2020matching,
  author  = {Xiumin Wang and Jianping Wang and Xinglin Zhang and
             Xiaoming Chen and Pan Zhou},
  title   = {Joint Task Offloading and Payment Determination for
             Mobile Edge Computing: A Stable Matching Based Approach},
  journal = {IEEE Transactions on Vehicular Technology},
  volume  = {69},
  number  = {10},
  pages   = {12148--12161},
  year    = {2020},
  doi     = {10.1109/TVT.2020.3013622}
}

@article{qi2024bridge,
  author  = {Houyi Qi and Minghui Liwang and Xianbin Wang and
             Li Li and Wei Gong and Jian Jin and Zhenzhen Jiao},
  title   = {Bridge the Present and Future: A Cross-Layer Matching Game
             in Dynamic Cloud-Aided Mobile Edge Networks},
  journal = {IEEE Transactions on Mobile Computing},
  volume  = {23},
  number  = {12},
  pages   = {12522--12539},
  year    = {2024},
  doi     = {10.1109/TMC.2024.3412751}
}

@article{amanatidis2025partial,
  author  = {Georgios Amanatidis and Sophie Klumper and
             Evangelos Markakis and Guido Sch{\"a}fer and
             Artem Tsikiridis},
  title   = {Partial Allocations in Budget-Feasible Mechanism Design:
             Bridging Multiple Levels of Service and Divisible Agents},
  journal = {ACM Transactions on Economics and Computation},
  volume  = {13},
  number  = {2},
  pages   = {1--28},
  year    = {2025},
  doi     = {10.1145/3723883}
}

@inproceedings{caragiannis2026divisible,
  author    = {Ioannis Caragiannis and Zhile Jiang and
               Apostolis Kerentzis},
  title     = {Bounds on the Revenue Gap of Linear Posted Pricing
               for Selling a Divisible Item},
  booktitle = {Web and Internet Economics: 21st International Conference,
               WINE 2025, Proceedings},
  series    = {Lecture Notes in Computer Science},
  volume    = {16266},
  pages     = {200--215},
  publisher = {Springer},
  address   = {Cham},
  year      = {2026},
  doi       = {10.1007/978-3-032-18660-7_10}
}

@article{chaisiri2012optimization,
  author  = {Sivadon Chaisiri and Bu-Sung Lee and Dusit Niyato},
  title   = {Optimization of Resource Provisioning Cost in Cloud Computing},
  journal = {IEEE Transactions on Services Computing},
  volume  = {5},
  number  = {2},
  pages   = {164--177},
  year    = {2012},
  doi     = {10.1109/TSC.2011.7}
}

@article{dazhi2025slicing,
  author  = {Michael N. Dazhi and Hayder Al-Hraishawi
             and M. R. Bhavani Shankar and Symeon Chatzinotas
             and Jo{\"e}l Grotz},
  title   = {Joint {NTN} Slicing and Admission Control
             for Infrastructure-as-a-Service: A Deep
             Learning Aided Multi-Objective Optimization},
  journal = {IEEE Transactions on Cognitive Communications
             and Networking},
  volume  = {11},
  number  = {2},
  pages   = {1297--1315},
  year    = {2025},
  doi     = {10.1109/TCCN.2024.3461673}
}

@inproceedings{barbarossa2017overbooking,
  author    = {Sergio Barbarossa and Elena Ceci and Mattia Merluzzi},
  title     = {Overbooking Radio and Computation Resources in {mmW}-Mobile Edge Computing to Reduce Vulnerability to Channel Intermittency},
  booktitle = {2017 European Conference on Networks and Communications (EuCNC)},
  pages     = {1--5},
  year      = {2017},
  publisher = {IEEE},
  doi       = {10.1109/EuCNC.2017.7980746}
}

@inproceedings{zhang2018dynamic,
  author    = {Fuming Zhang and Zhiqing Tang and Mingcheng Chen and Xiaojie Zhou and Weijia Jia},
  title     = {A Dynamic Resource Overbooking Mechanism in Fog Computing},
  booktitle = {2018 {IEEE} 15th International Conference on Mobile Ad Hoc and Sensor Systems ({MASS})},
  pages     = {89--97},
  year      = {2018},
  publisher = {IEEE},
  doi       = {10.1109/MASS.2018.00023}
}

@article{sexton2020provisioning,
  author  = {Conor Sexton and Nicola Marchetti and Luiz A. DaSilva},
  title   = {On Provisioning Slices and Overbooking Resources in Service Tailored Networks of the Future},
  journal = {IEEE/ACM Transactions on Networking},
  volume  = {28},
  number  = {5},
  pages   = {2106--2119},
  year    = {2020},
  doi     = {10.1109/TNET.2020.3004443}
}

@article{sheng2020futures,
  author  = {Shuran Sheng and Ruitao Chen and Peng Chen and Xianbin Wang and Lenan Wu},
  title   = {Futures-Based Resource Trading and Fair Pricing in Real-Time {IoT} Networks},
  journal = {IEEE Wireless Communications Letters},
  volume  = {9},
  number  = {1},
  pages   = {125--128},
  year    = {2020},
  doi     = {10.1109/LWC.2019.2945025}
}

@article{liwang2022overbooking,
  author  = {Minghui Liwang and Xianbin Wang},
  title   = {Overbooking-Empowered Computing Resource Provisioning in Cloud-Aided Mobile Edge Networks},
  journal = {IEEE/ACM Transactions on Networking},
  volume  = {30},
  number  = {5},
  pages   = {2289--2303},
  year    = {2022},
  doi     = {10.1109/TNET.2022.3167396}
}

@inproceedings{ma2026following,
  author    = {Tie Ma and Shan Zhang and Xiaoyu Zhang and Zichuan Zheng and Zhiyuan Wang and Hongbin Luo},
  title     = {Following the Usage, Not the Request: Risk-Aware Task Scheduling with Overbooking in Edge Clouds},
  booktitle = {IEEE INFOCOM 2026 - IEEE Conference on Computer Communications},
  pages     = {1--10},
  year      = {2026},
  publisher = {IEEE},
  doi       = {10.1109/INFOCOM59046.2026.11571458}
}

@inproceedings{vaswani2017attention,
  author    = {Ashish Vaswani and Noam Shazeer
               and Niki Parmar and Jakob Uszkoreit
               and Llion Jones and Aidan N. Gomez
               and {\L}ukasz Kaiser and Illia Polosukhin},
  title     = {Attention Is All You Need},
  booktitle = {Advances in Neural Information Processing Systems},
  volume    = {30},
  pages     = {5998--6008},
  year      = {2017}
}

@article{harsanyi1972generalized,
  author  = {John C. Harsanyi and Reinhard Selten},
  title   = {A Generalized Nash Solution for Two-Person
             Bargaining Games with Incomplete Information},
  journal = {Management Science},
  volume  = {18},
  number  = {5-part-2},
  pages   = {80--106},
  year    = {1972},
  doi     = {10.1287/mnsc.18.5.80}
}

@article{hatfield2005matching,
  author  = {John William Hatfield and Paul R. Milgrom},
  title   = {Matching with Contracts},
  journal = {American Economic Review},
  volume  = {95},
  number  = {4},
  pages   = {913--935},
  year    = {2005},
  doi     = {10.1257/0002828054825466}
}

@article{pisinger1995minimal,
  author  = {David Pisinger},
  title   = {A Minimal Algorithm for the
             Multiple-Choice Knapsack Problem},
  journal = {European Journal of Operational Research},
  volume  = {83},
  number  = {2},
  pages   = {394--410},
  year    = {1995},
  doi     = {10.1016/0377-2217(95)00015-I}
}

@inproceedings{sharafaldin2019developing,
  author    = {Iman Sharafaldin and Arash Habibi Lashkari and
               Saqib Hakak and Ali A. Ghorbani},
  title     = {Developing Realistic Distributed Denial of Service
               ({DDoS}) Attack Dataset and Taxonomy},
  booktitle = {2019 International Carnahan Conference on Security
               Technology (ICCST)},
  pages     = {1--8},
  year      = {2019},
  publisher = {IEEE},
  doi       = {10.1109/CCST.2019.8888419}
}

\clearpage
\appendices
\section{Supplementary Material}
\label{sec:supplementary_material}
\paragraph{Notation and scope of risk guarantees.}
Throughout this supplement, $\mathcal R^{\mathrm{true}}$ denotes the
probability-defined risk in the main text, and
$\widehat{\mathcal R}^{\mathrm{emp}}$ denotes its empirical
residual-scenario estimate computed from the fixed validation-residual
scenarios. All risk-feasibility results in this supplement are stated
with respect to $\widehat{\mathcal R}^{\mathrm{emp}}$. They are not
distribution-free guarantees on $\mathcal R^{\mathrm{true}}$. When no
ambiguity arises, we write $\mathcal R$ for
$\widehat{\mathcal R}^{\mathrm{emp}}$.

\subsection{Finite-Termination Analysis}
\label{app:finite_termination}

We first establish the finite termination of the contract-formation
and online-fulfillment procedures for a fixed target timeslot $t$.
The analysis concerns the staged contract mechanism after the
prediction inputs and the corresponding finite candidate-contract
sets have been constructed. In particular, it does not assert
convergence of the predictor-training procedure.
Throughout the final contract-adjustment procedure, once a BE--ES
pair is released or rejected, the corresponding ES remains recorded
in the BE's final rejection set for the remainder of the current
adjustment process, including any subsequent adjustment iterations
triggered by E\&C contract cancellation. Hence, such a pair is not
reconsidered within the same target timeslot.

\begin{lemma}[Finite termination of interim B\&E matching]
\label{lem:interim_matching_termination}
For any fixed target timeslot $t$, the interim B\&E
application--rejection procedure terminates after finitely many
applications.
\end{lemma}

\begin{proof}
For each BE $b_i\in\mathcal B$, let
$\mathcal L_i^{(t),0}$ denote its initial preference list.
By construction,
$
\mathcal L_i^{(t),0}\subseteq\mathcal S,
$
and therefore
\begin{equation}
\left|
\mathcal L_i^{(t),0}
\right|
\leq
|\mathcal S|.
\label{eq:interim_initial_list_bound}
\end{equation}
During interim matching, an unmatched BE applies to the
most-preferred ES remaining in its current working preference list.
Whenever $b_i$ is rejected by $s_j$, the ES $s_j$ is removed from
that working list and is not reinserted during the same interim
matching procedure. Consequently, each ordered pair
$(b_i,s_j)$ can generate at most one interim application.
Let
$N_{\mathrm{app}}^{\mathsf{int},(t)}$
denote the total number of interim BE-to-ES applications in
timeslot $t$. Then
\begin{equation}
N_{\mathrm{app}}^{\mathsf{int},(t)}
\leq
\sum_{b_i\in\mathcal B}
\left|
\mathcal L_i^{(t),0}
\right|
\leq
|\mathcal B|\,|\mathcal S|.
\label{eq:interim_application_bound}
\end{equation}
For every application event, the receiving ES performs its interim
subset-selection operation over a finite candidate-BE set. Under the
booked-rate discretization used by the interim selector, the
associated dynamic program has a finite number of discretized states
and therefore terminates after finitely many operations. This argument
concerns only finite termination of the discretized selector and does
not require global optimality with respect to the underlying
undiscretized combinatorial problem.
Since only finitely many applications can occur and every ES-side
selection invoked by an application terminates finitely, the interim
B\&E application--rejection procedure terminates.
\end{proof}

\begin{lemma}[Finite termination of supplementary reservation and final
B\&E contract adjustment]
\label{lem:final_adjustment_termination}
For any fixed target timeslot $t$, the E\&C
supplementary-reservation procedure, the final risk-aware B\&E
contract-adjustment procedure, and the associated
payment-feasibility loop terminate after finitely many operations.
\end{lemma}

\begin{proof}
We consider the CS-side supplementary-capacity allocation, the
risk-triggered B\&E adjustment, and the outer payment-feasibility
loop in turn.First, for each ES $s_j$, the feasible supplementary-reservation
interval
$
\left[
0,
\overline r_j^{S,(t)}
\right]
$
is discretized using the positive granularity $g_x>0$.
Hence, the number of candidate supplementary reserved rates is
finite and satisfies
\begin{equation}
K_j+1
=
\left\lceil
\frac{
\overline r_j^{S,(t)}
}{
g_x
}
\right\rceil
+1.
\label{eq:finite_cloud_candidate_levels}
\end{equation}
The CS-side multiple-choice dynamic program therefore operates over
a finite number of ES indices, discretized capacity states, and
candidate-level transitions. It consequently terminates after
finitely many operations.

Second, consider the final risk-aware B\&E contract adjustment for
a fixed supplementary-reservation outcome
$\{x_j^{E\leftrightarrow C,(t)}\}_{s_j\in\mathcal S}$.
For any risk-excess ES $s_j$, the mechanism sequentially releases
retained B\&E contracts until
$
\mathcal R_{j,\mathsf{sh}}^{S,(t)}
\left(
\mathcal B_j^{\mathsf{cur},(t)},
x_j^{E\leftrightarrow C,(t)}
\right)
\leq
\mathcal{Z}_j^S.
$
Each nonterminal release step removes at least one currently retained
B\&E contract from
$\mathcal B_j^{\mathsf{cur},(t)}$.
Moreover, the release procedure cannot continue indefinitely.
Indeed, if all B\&E contracts associated with $s_j$ were released,
then
$
\widetilde H_j^{(t)}(\emptyset)=0.
$
Under the empirical residual-scenario estimate, the aggregate
shortfall risk of the empty set satisfies
\begin{equation}
\widehat{\mathcal R}_{j,\mathsf{sh}}^{\mathrm{emp},(t)}
\left(
\emptyset,
x_j^{E\leftrightarrow C,(t)}
\right)
=
0.
\label{eq:empty_set_zero_shortfall_risk}
\end{equation}
Since $\mathcal{Z}_j^S\geq0$, the empty acceptance set is necessarily
final risk-feasible. Therefore, for a fixed current BE set, the
risk-triggered release procedure must terminate after at most the
number of currently retained B\&E contracts.
Each released BE may subsequently reapply to ESs in its final
admissible list. Whenever a BE is rejected during final
reapplication, the rejecting ES is removed from that BE's final
admissible list. Likewise, whenever an existing B\&E contract is
released because of aggregate shortfall risk, the releasing ES is
recorded in the corresponding BE's final rejection set. These
rejection records are preserved throughout the current final
adjustment process. Hence, once an ordered pair $(b_i,s_j)$ has been
released or rejected in the final stage, that same pair cannot be
reconsidered later in the same adjustment process.
Because there are only
$|\mathcal B||\mathcal S|$
possible BE--ES pairs, only finitely many final-stage applications,
rejections, and risk-triggered releases can occur.

Third, consider the outer payment-feasibility loop. Let
\begin{equation}
M^{E\leftrightarrow C,(t)}
=
\left|
\left\{
s_j\in\mathcal S
\;\middle|\;
x_j^{E\leftrightarrow C,(t)}>0
\right\}
\right|
\label{eq:active_ec_contract_count}
\end{equation}
denote the number of currently retained positive E\&C supplementary
contracts. In every nonterminal iteration of the payment-feasibility
loop, at least one retained payment-infeasible E\&C contract is
canceled by setting
$
x_j^{E\leftrightarrow C,(t)}=0,
p_j^{E\leftrightarrow C,(t)}=0.
$
The CS-side supplementary-capacity allocation is not rerun inside
this loop. Therefore, once an E\&C contract has been canceled, it
cannot be reconstructed during a later payment-feasibility iteration.
It follows that
$M^{E\leftrightarrow C,(t)}$
strictly decreases whenever the outer loop does not terminate.
Since each ES has at most one E\&C supplementary contract in
timeslot $t$,
\begin{equation}
0
\leq
M^{E\leftrightarrow C,(t)}
\leq
|\mathcal S|,
\end{equation}
and thus at most $|\mathcal S|$ E\&C contract-cancellation events
can occur.
An E\&C cancellation may trigger another risk-aware B\&E adjustment
from the current B\&E outcome. However, such an adjustment is finite
by the preceding argument, previously canceled E\&C contracts are
not restored, and previously released or rejected BE--ES pairs are
not reconsidered. Therefore, no adjustment cycle can repeat
indefinitely. Consequently, the combined E\&C supplementary-reservation, final
risk-aware B\&E contract-adjustment, and payment-feasibility
procedure terminates after finitely many operations.
\end{proof}

\begin{lemma}[Finite termination of online contract fulfillment]
\label{lem:online_fulfillment_termination}
For any fixed target timeslot $t$, the online contract-fulfillment
procedure terminates after finitely many weighted proportional
allocation rounds.
\end{lemma}

\begin{proof}
Consider one invocation of the weighted proportional allocation
operator
$
\operatorname{WPA}
\left(
\{u_i\}_{i\in\mathcal I},
C,
\{\mathcal{O}_i\}_{i\in\mathcal I}
\right),
$
where $\mathcal I$ is the finite set of currently active demands,
$u_i\geq0$ is the remaining demand of BE $b_i$, and $C\geq0$ is the
available capacity.
If
$
\sum_{i\in\mathcal I}u_i
\leq
C,
$
all remaining demands are fully satisfied and the invocation
terminates immediately. Otherwise, the operator computes
provisional proportional allocations.
If no provisional allocation exceeds the corresponding remaining
demand, the full remaining capacity is distributed among the active
demands and the invocation terminates. If at least one provisional
allocation exceeds its corresponding remaining demand, that
allocation is truncated to the remaining demand of the corresponding
BE. Such a BE is then fully satisfied and removed from the active
set before the next redistribution round.
Hence, whenever another redistribution round is required, the
cardinality of the active-demand set strictly decreases:
\begin{equation}
\left|
\mathcal I^{(r+1)}
\right|
<
\left|
\mathcal I^{(r)}
\right|,
\label{eq:wpa_active_set_decrease}
\end{equation}
where $\mathcal I^{(r)}$ denotes the active set in redistribution
round $r$.
The zero-total-weight case does not create a stalled state. When the
aggregate weight of the remaining active demands is zero, the
prescribed unweighted proportional fallback rule is used instead.
Therefore, an invocation starting with active set $\mathcal I$
requires at most
$
|\mathcal I|
$
redistribution rounds.

For each ES $s_j$, the online contract-fulfillment procedure invokes
WPA once for local edge-side fulfillment and at most once for
cloud-side supplementary fulfillment. Hence, the total number of WPA
redistribution rounds in timeslot $t$ is bounded by
\begin{equation}
N_{\mathrm{WPA}}^{(t)}
\leq
2
\sum_{s_j\in\mathcal S}
\left|
\mathcal F_j^{(t)}
\right|.
\label{eq:wpa_round_bound_general}
\end{equation}
Since each BE has at most one effective B\&E contract,
\begin{equation}
\sum_{s_j\in\mathcal S}
\left|
\mathcal F_j^{(t)}
\right|
\leq
|\mathcal B|,
\label{eq:effective_contract_partition_bound}
\end{equation}
and therefore
\begin{equation}
N_{\mathrm{WPA}}^{(t)}
\leq
2|\mathcal B|.
\label{eq:wpa_round_bound}
\end{equation}
Thus, the online contract-fulfillment procedure terminates after
finitely many operations.
\end{proof}

\begin{theorem}[Finite termination of contract formation and fulfillment]
\label{thm:finite_termination}
For every fixed target timeslot $t$, once the prediction inputs and
the corresponding finite candidate-contract sets have been
constructed, the proposed offline contract-formation procedure and
online contract-fulfillment procedure terminate after finitely many
operations.
\end{theorem}

\begin{proof}
The interim B\&E application--rejection procedure terminates after
finitely many operations by
Lemma~\ref{lem:interim_matching_termination}.
The E\&C supplementary-reservation procedure, final risk-aware B\&E
contract-adjustment procedure, and payment-feasibility loop terminate
after finitely many operations by
Lemma~\ref{lem:final_adjustment_termination}.
Finally, finite termination of the online contract-fulfillment
procedure follows from
Lemma~\ref{lem:online_fulfillment_termination}.
Therefore, for every fixed target timeslot $t$, both the offline
contract-formation stage and the online contract-fulfillment stage
terminate after finitely many operations.
\end{proof}

\subsection{Final Contract Feasibility}
\label{app:final_contract_feasibility}

We next establish the feasibility of the terminal offline
contract-formation outcome. The result concerns feasibility of the
formed contracts and resource commitments, rather than the
nonnegativity of the participants' ex-ante proxy objectives.
Ex-ante proxy rationality is analyzed separately.
For the B\&E pricing expressions below, we consider positive-volume
candidate contracts, i.e.,
$y_i^{B,\mathsf{book},(t)}>0$.
If the predicted demand of a BE is zero, no positive-volume B\&E
reservation contract is generated for that BE and the no-contract
option is retained.
Moreover, the terminal condition of the final-adjustment procedure is
understood to require simultaneously that
\begin{equation}
\mathcal S_{\mathsf{risk}}^{(t)}
=
\emptyset
\label{eq:terminal_no_risk_excess}
\end{equation}
and that every retained E\&C supplementary contract is
payment-feasible. In particular, whenever an E\&C contract is
canceled, the resulting shortfall risks are recomputed and the
risk-aware B\&E adjustment is repeated from the current B\&E
outcome before termination.

\begin{lemma}[BE-side feasibility of the final B\&E outcome]
\label{lem:be_final_feasibility}
For any fixed target timeslot $t$, every BE $b_i$ in the terminal
offline outcome is matched with at most one ES. Moreover, every
finally matched BE satisfies its reservation-risk requirement and
its B\&E reservation-payment budget.
\end{lemma}

\begin{proof}
By definition, the B\&E matching is a mapping
$
\mathcal{M}^{B\rightarrow E,(t)}
:
\mathcal B
\rightarrow
\mathcal S\cup\{\emptyset\}.
$
Hence, every BE is either matched with exactly one ES or remains
unmatched. Equivalently,
\begin{equation}
\sum_{s_j\in\mathcal S}
z_{i,j}^{(t)}
\leq
1,
\qquad
\forall b_i\in\mathcal B.
\label{eq:final_be_single_matching_feasibility}
\end{equation}
We next consider the reservation-risk requirement. A candidate ES is
included in the initial feasible set of BE $b_i$ only if
\begin{equation}
\mathcal R_i^{B,(t)}
\leq
\overline{\mathcal R}_i^B.
\label{eq:final_be_risk_feasibility}
\end{equation}
The interim application procedure considers only ESs in the
corresponding initial preference list
$\mathcal L_i^{(t),0}$.
During final adjustment, released or previously unmatched BEs reapply
only to ESs retained from this initial admissible domain, subject to
additional final-stage feasibility restrictions. Therefore, any ES
with which $b_i$ is finally matched was already admissible under the
BE-side reservation-risk requirement. Since the B\&E reserved rate
and its associated contract terms are not renegotiated during the
final adjustment, condition
\eqref{eq:final_be_risk_feasibility}
remains valid for every finally matched BE.
It remains to verify payment-budget feasibility. For a
positive-volume candidate B\&E contract, the negotiated reservation
price satisfies
\begin{equation}
p_{i,j}^{B\leftrightarrow E,*,(t)}
\leq
\frac{
\overline C_i^{B,(t)}
}{
y_i^{B,\mathsf{book},(t)}
}.
\label{eq:final_be_price_budget_bound}
\end{equation}
Because
$
y_i^{B,\mathsf{book},(t)}
=
x_{i,j}^{B\leftrightarrow E,(t)}
\Delta t,
$
the reservation payment associated with a finally retained B\&E
contract satisfies
\begin{align}
P_i^{B,(t)}
&=
p_{i,j}^{B\leftrightarrow E,*,(t)}
x_{i,j}^{B\leftrightarrow E,(t)}
\Delta t
\nonumber
\\
&=
p_{i,j}^{B\leftrightarrow E,*,(t)}
y_i^{B,\mathsf{book},(t)}
\nonumber
\\
&\leq
\overline C_i^{B,(t)}.
\label{eq:final_be_budget_feasibility}
\end{align}
The reserved rate and B\&E reservation price remain fixed during
subsequent E\&C reservation and final B\&E adjustment. Hence, later
contract-adjustment operations cannot invalidate this budget bound.
Therefore, the terminal B\&E outcome satisfies the single-matching,
BE reservation-risk, and BE payment-budget constraints.
\end{proof}

\begin{lemma}[ES-side feasibility of the final B\&E outcome]
\label{lem:es_final_feasibility}
For any fixed target timeslot $t$, the terminal BE set associated
with every ES $s_j$ satisfies the concurrency constraint, the maximum
controlled-overbooking constraint, and the final aggregate
shortfall-risk constraint.
\end{lemma}

\begin{proof}
Let
$
\mathcal B_j^{(t)}
=
\mathcal{M}^{E\leftarrow B,(t)}(s_j)
$
denote the terminal BE set of ES $s_j$.
During interim B\&E matching, every tentative acceptance set is
required to satisfy
\begin{equation}
\left|
\mathcal B_j^{(t)}
\right|
\leq
N_j^{\mathsf{max}}
\label{eq:final_es_concurrency}
\end{equation}
and
\begin{equation}
\sum_{b_i\in\mathcal B_j^{(t)}}
x_{i,j}^{B\leftrightarrow E,(t)}
\leq
\mathsf{OR}_j^{S,\mathsf{max}}f_j^S.
\label{eq:final_es_overbooking}
\end{equation}
After the formal E\&C supplementary reserved rate
$x_j^{E\leftrightarrow C,(t)}$
is determined, the final feasible family
$\mathfrak F_j^{S,\mathsf{fin},(t)}$
contains only BE sets satisfying
\eqref{eq:final_es_concurrency},
\eqref{eq:final_es_overbooking}, and
\begin{equation}
\mathcal R_{j,\mathsf{sh}}^{S,(t)}
\left(
\mathcal B_j^{(t)},
x_j^{E\leftrightarrow C,(t)}
\right)
\leq
\mathcal{Z}_j^S.
\label{eq:final_es_risk}
\end{equation}
The risk-triggered release operation only removes existing B\&E
contracts. Therefore, it cannot increase either the number of
retained BEs or the aggregate booked B\&E reservation rate, and hence
cannot violate
\eqref{eq:final_es_concurrency}
or
\eqref{eq:final_es_overbooking}.
When an affected BE reapplies during the final stage, an ES selects
its updated acceptance set only from
$\mathfrak F_j^{S,\mathsf{fin},(t)}$.
Consequently, every acceptance set produced by final reapplication
satisfies all three constraints.
An E\&C payment-feasibility cancellation may subsequently reduce
$x_j^{E\leftrightarrow C,(t)}$, which can increase the aggregate
shortfall risk of the corresponding ES. For this reason, the
mechanism recomputes
$
\mathcal R_{j,\mathsf{sh}}^{S,(t)}
\left(
\mathcal B_j^{(t)},
x_j^{E\leftrightarrow C,(t)}
\right)
$
after every such cancellation and repeats the risk-aware B\&E
adjustment from the current B\&E outcome whenever the risk threshold
is violated.
At termination,
$\mathcal S_{\mathsf{risk}}^{(t)}=\emptyset$.
Hence,
\begin{equation}
\mathcal R_{j,\mathsf{sh}}^{S,(t)}
\left(
\mathcal B_j^{(t)},
x_j^{E\leftrightarrow C,(t)}
\right)
\leq
\mathcal{Z}_j^S,
\qquad
\forall s_j\in\mathcal S.
\label{eq:terminal_es_risk_feasibility}
\end{equation}
Therefore, every terminal ES acceptance set satisfies the concurrency,
controlled-overbooking, and final aggregate shortfall-risk constraints.
\end{proof}

\begin{lemma}[CS-capacity and retained E\&C payment feasibility]
\label{lem:ec_final_feasibility}
For any fixed target timeslot $t$, the terminal E\&C
supplementary-contract assignment satisfies the CS physical
scrubbing-rate constraint, and every retained positive E\&C contract
is payment-feasible.
\end{lemma}

\begin{proof}
For each ES $s_j$, let $a_{j,k_j^*}$ denote the supplementary
reserved rate selected by the CS-side multiple-choice dynamic program.
The discretized capacity weight of candidate level $k$ is
$
w_{j,k}
=
\left\lceil
\frac{a_{j,k}}{g_x}
\right\rceil,
$
and the discretized CS capacity is
$
W^C
=
\left\lfloor
\frac{f^C}{g_x}
\right\rfloor.
$
The CS-side dynamic program admits only selections satisfying
\begin{equation}
\sum_{s_j\in\mathcal S}
w_{j,k_j^*}
\leq
W^C.
\label{eq:discrete_cloud_capacity}
\end{equation}
By the definition of the ceiling operator,
$
a_{j,k_j^*}
\leq
g_x
w_{j,k_j^*}.
$
Therefore,
\begin{align}
\sum_{s_j\in\mathcal S}
x_j^{E\leftrightarrow C,(t)}
&=
\sum_{s_j\in\mathcal S}
a_{j,k_j^*}
\nonumber
\\
&\leq
g_x
\sum_{s_j\in\mathcal S}
w_{j,k_j^*}
\nonumber
\\
&\leq
g_x W^C
\nonumber
\\
&\leq
f^C.
\label{eq:terminal_cloud_capacity_feasibility}
\end{align}
Thus, the initially formed E\&C supplementary-contract assignment
satisfies the physical scrubbing-rate limit of the CS.
Subsequent payment-feasibility operations can only cancel an existing
E\&C contract by replacing
$x_j^{E\leftrightarrow C,(t)}$
with zero. They never increase any supplementary reserved rate and do
not rerun the CS-side allocation. Hence,
\eqref{eq:terminal_cloud_capacity_feasibility}
continues to hold for the terminal E\&C assignment.
Finally, after each risk-aware B\&E adjustment, the mechanism
recomputes the current B\&E reservation revenue
$I_j^{S,\mathsf{fin},(t)}$ of every ES retaining a positive E\&C
contract and rechecks the cloud-side payment-feasibility condition
\begin{equation}
p_j^{E\leftrightarrow C,(t)}
x_j^{E\leftrightarrow C,(t)}
\Delta t
\leq
\mathcal{G}_j^S I_j^{S,\mathsf{fin},(t)}.
\label{eq:final_ec_payment_feasibility}
\end{equation}
Any E\&C contract that violates
\eqref{eq:final_ec_payment_feasibility} is canceled. The outer
adjustment procedure terminates only when every retained E\&C
contract is payment-feasible.
Therefore, the terminal E\&C assignment satisfies the CS physical
capacity constraint and contains no payment-infeasible retained
E\&C contract.
\end{proof}

\begin{theorem}[Final contract feasibility]
\label{thm:final_contract_feasibility}
For any fixed target timeslot $t$, the terminal offline
contract-formation outcome returned by the proposed mechanism is
contract-feasible. Specifically:

\begin{enumerate}
    \item every BE is matched with at most one ES, and every finally
    matched BE satisfies its reservation-risk requirement and B\&E
    reservation-payment budget;

    \item every ES satisfies its concurrency limit, maximum
    controlled-overbooking constraint, and final aggregate
    shortfall-risk constraint;

    \item the retained E\&C supplementary reservations satisfy the
    physical scrubbing-rate limit of the CS; and

    \item every retained positive E\&C supplementary contract is
    payment-feasible.
\end{enumerate}
\end{theorem}

\begin{proof}
Lemma~\ref{lem:be_final_feasibility} establishes the BE-side
single-matching, reservation-risk, and payment-budget constraints.
Lemma~\ref{lem:es_final_feasibility} establishes the ES-side
concurrency, controlled-overbooking, and aggregate shortfall-risk
constraints in the terminal B\&E outcome.
Lemma~\ref{lem:ec_final_feasibility} establishes the CS physical
capacity constraint and the payment feasibility of every retained
E\&C supplementary contract.
Combining these results yields the stated final contract-feasibility
property.
\end{proof}

\begin{remark}[Scope of final contract feasibility]
\label{rem:final_feasibility_scope}
Final contract feasibility does not require the aggregate B\&E
reserved rate of each ES to be deterministically covered by its
combined local and cloud-side physical capacities. Controlled
overbooking is explicitly permitted. The final B\&E outcome instead
satisfies the prescribed maximum overbooking bound together with the
aggregate shortfall-risk constraint in
\eqref{eq:terminal_es_risk_feasibility}.
Accordingly, Theorem~\ref{thm:final_contract_feasibility} is a
contract- and risk-feasibility result, not a deterministic
full-coverage guarantee.
\end{remark}

\subsection{Ex-Ante Proxy Rationality}
\label{app:ex_ante_participation}

We next characterize the proxy-rationality properties of the
offline contract-formation mechanism. These properties are stated in
terms of the ex-ante proxy objectives and bilateral bargaining
surpluses available before traffic realization. They are distinct
from the contract-feasibility properties established in
Theorem~\ref{thm:final_contract_feasibility} and do not constitute
guarantees on realized ex-post utilities.
The no-contract outside option is normalized to zero throughout this
subsection.

\begin{lemma}[BE-side ex-ante proxy rationality]
\label{lem:be_ex_ante_ir}
For any fixed target timeslot $t$, every finally matched BE
$b_i$ obtains a nonnegative BE-side ex-ante proxy objective:
\begin{equation}
\Pi_{i,\mathcal{M}^{B\rightarrow E,(t)}(b_i)}^{B,(t)}
\geq
0.
\label{eq:terminal_be_proxy_nonnegative}
\end{equation}
\end{lemma}

\begin{proof}
Consider any positive-volume price-feasible B\&E candidate pair
$(b_i,s_j)$. Its feasible unit-price interval satisfies
\begin{equation}
p_{i,j}^{B\leftrightarrow E,*,(t)}
\leq
\frac{
V_{i,j}^{B,\mathsf{pred},(t)}
}{
y_i^{B,\mathsf{book},(t)}
}.
\label{eq:be_ir_price_upper_bound}
\end{equation}
Hence,
\begin{align}
\Pi_{i,j}^{B,(t)}
&=
V_{i,j}^{B,\mathsf{pred},(t)}
-
p_{i,j}^{B\leftrightarrow E,*,(t)}
y_i^{B,\mathsf{book},(t)}
\nonumber
\\
&\geq
0.
\label{eq:be_proxy_nonnegative_from_price}
\end{align}
Moreover, the initial preference list of BE $b_i$ is constructed as
$
\mathcal L_i^{(t),0}
=
\left\{
s_j\in\mathcal S_i^{B,\mathsf{feas},(t)}
\;\middle|\;
\Pi_{i,j}^{B,(t)}\geq0
\right\}.
$
Therefore, the interim application process only considers B\&E
contracts having nonnegative BE-side proxy value.
During the final adjustment, an affected BE may reapply only to ESs
preserved from its initial preference list, after removing the ESs in
its final rejection set. The B\&E reserved rate and negotiated unit
reservation price are not renegotiated during this stage. Hence, any
B\&E contract retained in the terminal outcome preserves the same
nonnegative BE-side proxy value that qualified it for the initial
preference list.
Therefore, every finally matched BE satisfies
\eqref{eq:terminal_be_proxy_nonnegative}. An unmatched BE receives the
normalized no-contract proxy value zero.
\end{proof}

\begin{lemma}[Bilateral participation rationality of B\&E bargaining]
\label{lem:be_bilateral_participation}
For every positive-volume price-feasible B\&E candidate contract
$(b_i,s_j)$, the negotiated unit reservation price yields
nonnegative bilateral bargaining surpluses for both the BE and the ES:
\begin{equation}
S_{i,j}^{B,(t)}
\left(
p_{i,j}^{B\leftrightarrow E,*,(t)}
\right)
\geq0
\label{eq:be_bilateral_be_surplus_nonnegative}
\end{equation}
and
\begin{equation}
S_{i,j}^{S,(t)}
\left(
p_{i,j}^{B\leftrightarrow E,*,(t)}
\right)
\geq0.
\label{eq:be_bilateral_es_surplus_nonnegative}
\end{equation}
\end{lemma}

\begin{proof}
For a positive-volume B\&E candidate contract, the feasible
unit-price interval is
\begin{equation}
\underline p_{i,j}^{B\leftrightarrow E,(t)}
=
c_j^{S,(t)}
\leq
p
\leq
\overline p_{i,j}^{B\leftrightarrow E,(t)},
\end{equation}
where
\begin{equation}
\overline p_{i,j}^{B\leftrightarrow E,(t)}
=
\min
\left\{
\frac{
V_{i,j}^{B,\mathsf{pred},(t)}
}{
y_i^{B,\mathsf{book},(t)}
},
\frac{
\overline C_i^{B,(t)}
}{
y_i^{B,\mathsf{book},(t)}
}
\right\}.
\end{equation}

Price feasibility implies
$
c_j^{S,(t)}
\leq
\frac{
V_{i,j}^{B,\mathsf{pred},(t)}
}{
y_i^{B,\mathsf{book},(t)}
}
$
and
$
c_j^{S,(t)}
\leq
\frac{
\overline C_i^{B,(t)}
}{
y_i^{B,\mathsf{book},(t)}
}.
$
The unconstrained Nash-bargaining price
$
p_{i,j}^{B\leftrightarrow E,\mathsf{NB},(t)}
=
(1-\mathcal{W}_{i,j}^B)
\frac{
V_{i,j}^{B,\mathsf{pred},(t)}
}{
y_i^{B,\mathsf{book},(t)}
}
+
\mathcal{W}_{i,j}^B c_j^{S,(t)}
$
lies between
$c_j^{S,(t)}$
and
$V_{i,j}^{B,\mathsf{pred},(t)}
/
y_i^{B,\mathsf{book},(t)}$.
The implemented price
$
p_{i,j}^{B\leftrightarrow E,*,(t)}
=
\min
\left\{
p_{i,j}^{B\leftrightarrow E,\mathsf{NB},(t)},
\frac{
\overline C_i^{B,(t)}
}{
y_i^{B,\mathsf{book},(t)}
}
\right\}
$
therefore satisfies
\begin{equation}
c_j^{S,(t)}
\leq
p_{i,j}^{B\leftrightarrow E,*,(t)}
\leq
\frac{
V_{i,j}^{B,\mathsf{pred},(t)}
}{
y_i^{B,\mathsf{book},(t)}
}.
\label{eq:be_bargaining_price_interval_final}
\end{equation}
Using the BE-side bilateral surplus
$
S_{i,j}^{B,(t)}(p)
=
V_{i,j}^{B,\mathsf{pred},(t)}
-
p\,y_i^{B,\mathsf{book},(t)},
$
the upper bound in
\eqref{eq:be_bargaining_price_interval_final}
implies
$
S_{i,j}^{B,(t)}
\left(
p_{i,j}^{B\leftrightarrow E,*,(t)}
\right)
\geq0.
$
Similarly, because
$
C_{i,j}^{S,(t)}
=
c_j^{S,(t)}
y_i^{B,\mathsf{book},(t)},
$
the ES-side bilateral surplus is
$
S_{i,j}^{S,(t)}(p)
=
\left(
p-c_j^{S,(t)}
\right)
y_i^{B,\mathsf{book},(t)}.
$
The lower bound in
\eqref{eq:be_bargaining_price_interval_final}
therefore yields
$
S_{i,j}^{S,(t)}
\left(
p_{i,j}^{B\leftrightarrow E,*,(t)}
\right)
\geq0.
$
Thus, every price-feasible B\&E bargaining outcome is individually
acceptable to both bargaining parties relative to their bilateral
disagreement points.
\end{proof}

\begin{lemma}[Bilateral participation rationality of E\&C bargaining]
\label{lem:ec_bilateral_participation}
For every positive price-feasible E\&C candidate level
$a_{j,k}>0$, the negotiated E\&C unit reservation price yields
nonnegative bilateral bargaining surpluses for both ES $s_j$ and
the CS:
\begin{equation}
S_{j,k}^{S,E\leftrightarrow C,(t)}
\left(
p_{j,k}^{E\leftrightarrow C,*,(t)}
\right)
\geq0
\label{eq:ec_bilateral_es_surplus_nonnegative}
\end{equation}
and
\begin{equation}
S_{j,k}^{C,E\leftrightarrow C,(t)}
\left(
p_{j,k}^{E\leftrightarrow C,*,(t)}
\right)
\geq0.
\label{eq:ec_bilateral_cs_surplus_nonnegative}
\end{equation}
\end{lemma}

\begin{proof}
For a positive E\&C candidate level, let
$
A_{j,k}^{(t)}
=
a_{j,k}\Delta t
>
0.
$
The feasible E\&C unit-price interval is
\begin{equation}
c^{C,(t)}
=
\underline p_{j,k}^{E\leftrightarrow C,(t)}
\leq
p
\leq
\overline p_{j,k}^{E\leftrightarrow C,(t)},
\end{equation}
where
\begin{equation}
\overline p_{j,k}^{E\leftrightarrow C,(t)}
=
\min
\left\{
\frac{
V_{j,k}^{E\leftrightarrow C,(t)}
}{
A_{j,k}^{(t)}
},
\frac{
B_j^{S,\mathsf{cld},(t)}
}{
A_{j,k}^{(t)}
}
\right\}.
\end{equation}
Price feasibility therefore implies
$
c^{C,(t)}
\leq
\frac{
V_{j,k}^{E\leftrightarrow C,(t)}
}{
A_{j,k}^{(t)}
}
$
and
$
c^{C,(t)}
\leq
\frac{
B_j^{S,\mathsf{cld},(t)}
}{
A_{j,k}^{(t)}
}.
$
The unconstrained Nash-bargaining price
$
p_{j,k}^{E\leftrightarrow C,\mathsf{NB},(t)}
=
(1-\mathcal{W}_j^{EC})
\frac{
V_{j,k}^{E\leftrightarrow C,(t)}
}{
A_{j,k}^{(t)}
}
+
\mathcal{W}_j^{EC}c^{C,(t)}
$
lies between
$c^{C,(t)}$
and
$V_{j,k}^{E\leftrightarrow C,(t)}
/
A_{j,k}^{(t)}$.
After applying the cloud-side reservation-budget cap,
$
p_{j,k}^{E\leftrightarrow C,*,(t)}
=
\min
\left\{
p_{j,k}^{E\leftrightarrow C,\mathsf{NB},(t)},
\frac{
B_j^{S,\mathsf{cld},(t)}
}{
A_{j,k}^{(t)}
}
\right\},
$
we still have
\begin{equation}
c^{C,(t)}
\leq
p_{j,k}^{E\leftrightarrow C,*,(t)}
\leq
\frac{
V_{j,k}^{E\leftrightarrow C,(t)}
}{
A_{j,k}^{(t)}
}.
\label{eq:ec_bargaining_price_interval_final}
\end{equation}
The ES-side E\&C bargaining surplus is
$
S_{j,k}^{S,E\leftrightarrow C,(t)}(p)
=
V_{j,k}^{E\leftrightarrow C,(t)}
-
pA_{j,k}^{(t)},
$
and hence the upper bound in
\eqref{eq:ec_bargaining_price_interval_final}
gives
$
S_{j,k}^{S,E\leftrightarrow C,(t)}
\left(
p_{j,k}^{E\leftrightarrow C,*,(t)}
\right)
\geq0.
$
The CS-side E\&C bargaining surplus is
$
S_{j,k}^{C,E\leftrightarrow C,(t)}(p)
=
\left(
p-c^{C,(t)}
\right)
A_{j,k}^{(t)}.
$
The lower bound in
\eqref{eq:ec_bargaining_price_interval_final}
therefore gives
$
S_{j,k}^{C,E\leftrightarrow C,(t)}
\left(
p_{j,k}^{E\leftrightarrow C,*,(t)}
\right)
\geq0.
$
Thus, every positive price-feasible E\&C bargaining outcome is
bilaterally participation-rational.
\end{proof}

\begin{lemma}[CS-side ex-ante proxy rationality]
\label{lem:cs_ex_ante_ir}
For any fixed target timeslot $t$, the terminal E\&C assignment yields
a nonnegative aggregate CS-side ex-ante proxy value.
\end{lemma}

\begin{proof}
For every positive price-feasible E\&C candidate level,
Lemma~\ref{lem:ec_bilateral_participation} implies
$
p_{j,k}^{E\leftrightarrow C,*,(t)}
\geq
c^{C,(t)}.
$
Hence, its CS-side ex-ante proxy value satisfies
\begin{align}
G_{j,k}^{C,(t)}
&=
\left(
p_{j,k}^{E\leftrightarrow C,*,(t)}
-
c^{C,(t)}
\right)
a_{j,k}\Delta t
\nonumber
\\
&\geq
0.
\label{eq:cs_candidate_proxy_nonnegative}
\end{align}
For the zero-allocation candidate,
$
a_{j,0}=0,
\qquad
p_{j,0}^{E\leftrightarrow C,*,(t)}=0,
$
and therefore
$
G_{j,0}^{C,(t)}=0.
$
Let $\bar k_j^{(t)}$ denote the terminal E\&C level associated with
ES $s_j$, where
$\bar k_j^{(t)}=0$
if the originally selected positive E\&C contract is subsequently
canceled by the payment-feasibility procedure. Every retained
positive level has nonnegative CS-side proxy value by
\eqref{eq:cs_candidate_proxy_nonnegative}, while every canceled
contract contributes the zero-level proxy value.
Consequently,
\begin{equation}
\sum_{s_j\in\mathcal S}
G_{j,\bar k_j^{(t)}}^{C,(t)}
\geq
0.
\label{eq:terminal_cs_proxy_nonnegative}
\end{equation}
Thus, the terminal E\&C assignment is ex-ante proxy-rational
for the CS relative to the zero-allocation outside option.
\end{proof}

\begin{proposition}[Conditional ES-side ex-ante proxy rationality]
\label{prop:conditional_es_ex_ante_ir}
Fix a target timeslot $t$ and an ES $s_j$. Suppose that, with respect
to the terminal formal E\&C rate--price pair
$
\left(
x_j^{E\leftrightarrow C,(t)},
p_j^{E\leftrightarrow C,(t)}
\right),
$
the terminal accepted BE set
$\mathcal B_j^{\mathsf{fin},(t)}$
is selected by the final ES-side choice operator:
\begin{equation}
\mathcal B_j^{\mathsf{fin},(t)}
\in
\arg\max_{
\substack{
\mathcal B_j
\subseteq
\mathcal B_j^{\mathsf{cand,fin},(t)},\\
\mathcal B_j
\in
\mathfrak F_j^{S,\mathsf{fin},(t)}
}
}
\Pi_j^{S,\mathsf{upd},(t)}
\left(
\mathcal B_j
\right).
\label{eq:conditional_es_final_argmax}
\end{equation}
Then
\begin{equation}
\Pi_j^{S,\mathsf{upd},(t)}
\left(
\mathcal B_j^{\mathsf{fin},(t)}
\right)
\geq
0.
\label{eq:conditional_es_proxy_nonnegative}
\end{equation}
\end{proposition}

\begin{proof}
By construction, the empty set belongs to the final risk-feasible
family:
$
\emptyset
\in
\mathfrak F_j^{S,\mathsf{fin},(t)},
$
and the no-contract proxy value is normalized as
$
\Pi_j^{S,\mathsf{upd},(t)}(\emptyset)=0.
$
Because
$\emptyset\subseteq
\mathcal B_j^{\mathsf{cand,fin},(t)}$,
the empty set is feasible for the maximization problem in
\eqref{eq:conditional_es_final_argmax}. Therefore, optimality of
$\mathcal B_j^{\mathsf{fin},(t)}$ implies
\begin{align}
\Pi_j^{S,\mathsf{upd},(t)}
\left(
\mathcal B_j^{\mathsf{fin},(t)}
\right)
&\geq
\Pi_j^{S,\mathsf{upd},(t)}
\left(
\emptyset
\right)
\nonumber
\\
&=
0.
\end{align}
This proves the stated conditional ES-side ex-ante
proxy-rationality property.
\end{proof}

\begin{remark}[Scope of the participation-rationality guarantees]
\label{rem:participation_rationality_scope}
The preceding results distinguish bilateral participation rationality
from aggregate ES-side proxy rationality.
Lemmas~\ref{lem:be_ex_ante_ir}--\ref{lem:cs_ex_ante_ir} provide
unconditional guarantees for finally matched BEs, price-feasible
B\&E and E\&C bargaining outcomes, and the terminal CS-side proxy
value. In contrast, Proposition~\ref{prop:conditional_es_ex_ante_ir}
is conditional because the localized final-adjustment mechanism does
not necessarily re-optimize the updated aggregate ES-side proxy
objective for every ES in every terminal state.
In particular, final risk feasibility and E\&C payment feasibility
alone do not imply
$
\Pi_j^{S,\mathsf{upd},(t)}
\left(
\mathcal B_j^{(t)}
\right)
\geq0,
$
because the updated ES-side proxy additionally includes estimated
local operating cost, formal cloud-reservation payment, and the
aggregate shortfall-risk penalty.
Accordingly, the proposed mechanism does not claim unconditional
aggregate ex-ante proxy rationality for every terminal ES
outcome. It also does not claim incentive compatibility,
strategy-proofness, or truthfulness. All guarantees in this
subsection concern the specified offline ex-ante proxy objectives and
bilateral bargaining surpluses rather than realized ex-post utilities.
\end{remark}

\subsection{Conditional Restricted Stability}
\label{app:conditional_restricted_stability}

We finally analyze the no-blocking property induced by the localized
final B\&E reapplication process. Because the proposed mechanism does
not reopen all B\&E pairs after the formal E\&C reservation has been
determined, the resulting guarantee is intentionally restricted to
the BE--ES pairs actually explored by the terminal final-reapplication
process.
Consider the last final-adjustment iteration after the terminal
E\&C rate--price pairs
$
\left\{
\left(
x_j^{E\leftrightarrow C,(t)},
p_j^{E\leftrightarrow C,(t)}
\right)
\right\}_{s_j\in\mathcal S}
$
have become fixed, i.e., no subsequent E\&C cancellation occurs.
All ES-side proxy objectives and final risk-feasible families in this
subsection are evaluated with respect to these terminal E\&C terms.

\begin{definition}[Terminal ES choice function]
\label{def:terminal_es_choice}
For ES $s_j$, let
$\mathfrak F_j^{S,\star,(t)}$
denote its final risk-feasible family under the terminal E\&C
reservation, evaluated over the discretized candidate sets. For any
finite BE offer set $\mathcal A\subseteq\mathcal B$, define
\begin{equation}
\operatorname{Ch}_j^{S,\star,(t)}
\left(
\mathcal A
\right)
\in
\arg\max_{
\substack{
\mathcal B_j\subseteq\mathcal A,\\
\mathcal B_j\in
\mathfrak F_j^{S,\star,(t)}
}
}
\Pi_j^{S,\mathsf{upd},(t)}
\left(
\mathcal B_j
\right).
\label{eq:terminal_es_choice_function}
\end{equation}
The deterministic revenue-based and lexicographic tie-breaking rules
used by the mechanism make
$\operatorname{Ch}_j^{S,\star,(t)}$
single-valued.
\end{definition}

\begin{definition}[Terminal explored final-reapplication submarket]
\label{def:terminal_explored_submarket}
Let
$\mathcal S_{\mathsf{act}}^{(t)}$
denote the set of ESs for which the terminal ES choice function in
\eqref{eq:terminal_es_choice_function} is invoked at least once during
the last final-reapplication iteration.
For every
$s_j\in\mathcal S_{\mathsf{act}}^{(t)}$,
let
$\mathcal A_j^{\star,(t)}$
denote the cumulative set of BEs that are presented to
$\operatorname{Ch}_j^{S,\star,(t)}$
during that terminal iteration, including the BEs retained by $s_j$
when its first terminal choice is invoked and all BEs subsequently
applying to $s_j$.
The collection
$
\left\{
\left(
b_i,s_j
\right):
s_j\in\mathcal S_{\mathsf{act}}^{(t)},
\;
b_i\in\mathcal A_j^{\star,(t)}
\right\}
$
is called the \emph{terminal explored final-reapplication submarket}.
\end{definition}

\begin{definition}[Restricted final-reapplication blocking pair]
\label{def:restricted_final_blocking_pair}
Let
$\mathcal{M}^{B\rightarrow E,\star,(t)}$
denote the terminal B\&E matching, and let
$
\mathcal B_j^{\star,(t)}
=
\mathcal{M}^{E\leftarrow B,\star,(t)}(s_j)
$
denote the corresponding terminal BE set of ES $s_j$.
A pair $(b_i,s_j)$ is a
\emph{restricted final-reapplication blocking pair} if

\begin{enumerate}
    \item
    $s_j\in\mathcal S_{\mathsf{act}}^{(t)}$,
    $b_i\in\mathcal A_j^{\star,(t)}$, and
    $\mathcal{M}^{B\rightarrow E,\star,(t)}(b_i)\neq s_j$;

    \item
    BE $b_i$ strictly prefers $s_j$ to its terminal assignment:
    \begin{equation}
    s_j
    \succ_i^B
    \mathcal{M}^{B\rightarrow E,\star,(t)}(b_i);
    \label{eq:restricted_block_be_preference}
    \end{equation}

    \item
    if $b_i$ were added to the terminal BE set of $s_j$, then the
    terminal ES choice function would select a set containing $b_i$:
    \begin{equation}
    b_i
    \in
    \operatorname{Ch}_j^{S,\star,(t)}
    \left(
    \mathcal B_j^{\star,(t)}
    \cup
    \{b_i\}
    \right).
    \label{eq:restricted_block_es_choice}
    \end{equation}
\end{enumerate}

Condition~\eqref{eq:restricted_block_es_choice} allows the ES choice
function to drop one or more currently retained BEs and therefore
covers both expansion- and replacement-type deviations within the
explored submarket.
\end{definition}

\begin{lemma}[Irrelevance of rejected contracts for the terminal ES choice]
\label{lem:terminal_choice_irc}
For every ES $s_j$, the terminal choice function satisfies the
irrelevance-of-rejected-contracts property on its terminal domain.
Specifically, for any
$\mathcal A'\subseteq\mathcal A$ such that
\begin{equation}
\operatorname{Ch}_j^{S,\star,(t)}(\mathcal A)
\subseteq
\mathcal A',
\label{eq:irc_containment}
\end{equation}
we have
\begin{equation}
\operatorname{Ch}_j^{S,\star,(t)}(\mathcal A')
=
\operatorname{Ch}_j^{S,\star,(t)}(\mathcal A).
\label{eq:terminal_choice_irc}
\end{equation}
\end{lemma}

\begin{proof}
Let
$
\mathcal B^\star
=
\operatorname{Ch}_j^{S,\star,(t)}(\mathcal A).
$
By
\eqref{eq:irc_containment},
$\mathcal B^\star\subseteq\mathcal A'$,
so $\mathcal B^\star$ remains an admissible candidate when the offer
set is reduced from $\mathcal A$ to $\mathcal A'$.
Every final risk-feasible subset of $\mathcal A'$ is also a final
risk-feasible subset of $\mathcal A$, because
$\mathcal A'\subseteq\mathcal A$
and the terminal feasible family
$\mathfrak F_j^{S,\star,(t)}$
is fixed under the terminal E\&C assignment.
Since $\mathcal B^\star$ maximizes the same updated ES-side proxy
objective over the larger domain $\mathcal A$, no subset available in
$\mathcal A'$ can be strictly preferred to $\mathcal B^\star$.
The deterministic tie-breaking rule is independent of the surrounding
offer set and therefore selects the same maximizer if several subsets
have the same proxy value.
Hence,
$
\operatorname{Ch}_j^{S,\star,(t)}(\mathcal A')
=
\mathcal B^\star
=
\operatorname{Ch}_j^{S,\star,(t)}(\mathcal A).
$
\end{proof}

\begin{assumption}[Substitutability on the terminal explored domain]
\label{ass:terminal_choice_substitutability}
For every
$s_j\in\mathcal S_{\mathsf{act}}^{(t)}$,
the terminal ES choice function is substitutable over subsets of
$\mathcal A_j^{\star,(t)}$.
That is, for any
$
\mathcal A'
\subseteq
\mathcal A
\subseteq
\mathcal A_j^{\star,(t)}
$
and any $b_i\in\mathcal A'$,
\begin{equation}
b_i
\in
\operatorname{Ch}_j^{S,\star,(t)}
\left(
\mathcal A
\right)
\quad
\Longrightarrow
\quad
b_i
\in
\operatorname{Ch}_j^{S,\star,(t)}
\left(
\mathcal A'
\right).
\label{eq:terminal_substitutability}
\end{equation}
\end{assumption}

Assumption~\ref{ass:terminal_choice_substitutability} is an explicit
regularity condition for the restricted stability result. It is not
claimed to hold automatically for the general risk-coupled ES choice
problem, which includes heterogeneous reservation rates, concurrency
and controlled-overbooking constraints, and aggregate shortfall-risk
effects.

\begin{lemma}[Path independence on the explored terminal domain]
\label{lem:terminal_choice_path_independence}
Under
Assumption~\ref{ass:terminal_choice_substitutability},
the terminal ES choice function is path independent over the explored
terminal domain. In particular, for admissible offer sets
$\mathcal A$ and $\mathcal D$,
\begin{equation}
\operatorname{Ch}_j^{S,\star,(t)}
\left(
\operatorname{Ch}_j^{S,\star,(t)}
(\mathcal A)
\cup
\mathcal D
\right)
=
\operatorname{Ch}_j^{S,\star,(t)}
\left(
\mathcal A\cup\mathcal D
\right).
\label{eq:terminal_choice_path_independence}
\end{equation}
\end{lemma}

\begin{proof}
Let
$
\mathcal C
=
\operatorname{Ch}_j^{S,\star,(t)}
\left(
\mathcal A\cup\mathcal D
\right).
$
By substitutability,
every BE in
$\mathcal C\cap\mathcal A$
must also belong to
$\operatorname{Ch}_j^{S,\star,(t)}(\mathcal A)$.
Every BE in
$\mathcal C\setminus\mathcal A$
belongs to $\mathcal D$.
Therefore,
\begin{equation}
\mathcal C
\subseteq
\operatorname{Ch}_j^{S,\star,(t)}
(\mathcal A)
\cup
\mathcal D.
\label{eq:path_independence_containment}
\end{equation}
Moreover,
$
\operatorname{Ch}_j^{S,\star,(t)}
(\mathcal A)
\cup
\mathcal D
\subseteq
\mathcal A\cup\mathcal D.
$
Applying
Lemma~\ref{lem:terminal_choice_irc}
to
\eqref{eq:path_independence_containment}
gives
$
\operatorname{Ch}_j^{S,\star,(t)}
\left(
\operatorname{Ch}_j^{S,\star,(t)}
(\mathcal A)
\cup
\mathcal D
\right)
=
\mathcal C,
$
which proves
\eqref{eq:terminal_choice_path_independence}.
\end{proof}

\begin{lemma}[Consistency with cumulative terminal offers]
\label{lem:cumulative_terminal_offer_consistency}
Under
Assumption~\ref{ass:terminal_choice_substitutability},
for every activated ES $s_j$, its terminal retained BE set equals its
choice from the cumulative set of BEs explored during the terminal
reapplication process:
\begin{equation}
\mathcal B_j^{\star,(t)}
=
\operatorname{Ch}_j^{S,\star,(t)}
\left(
\mathcal A_j^{\star,(t)}
\right).
\label{eq:cumulative_terminal_choice}
\end{equation}
\end{lemma}

\begin{proof}
Consider the successive invocations of the terminal ES choice function
for a fixed activated ES $s_j$.
At its first invocation, let
$\mathcal A_{j,1}^{(t)}$
denote the candidate set consisting of the BEs currently retained by
$s_j$ together with the newly applying BEs. The mechanism selects
$
\mathcal H_{j,1}^{(t)}
=
\operatorname{Ch}_j^{S,\star,(t)}
\left(
\mathcal A_{j,1}^{(t)}
\right).
$
Suppose after invocation $r-1$,
\begin{equation}
\mathcal H_{j,r-1}^{(t)}
=
\operatorname{Ch}_j^{S,\star,(t)}
\left(
\mathcal A_{j,r-1}^{(t)}
\right),
\label{eq:cumulative_induction_hypothesis}
\end{equation}
where
$\mathcal A_{j,r-1}^{(t)}$
is the cumulative set of all BEs presented to $s_j$ up to that
invocation.
Let
$\mathcal N_{j,r}^{(t)}$
be the set of newly applying BEs at invocation $r$. The mechanism
updates its retained set as
\begin{align}
\mathcal H_{j,r}^{(t)}
&=
\operatorname{Ch}_j^{S,\star,(t)}
\left(
\mathcal H_{j,r-1}^{(t)}
\cup
\mathcal N_{j,r}^{(t)}
\right)
\nonumber
\\
&=
\operatorname{Ch}_j^{S,\star,(t)}
\left(
\operatorname{Ch}_j^{S,\star,(t)}
(
\mathcal A_{j,r-1}^{(t)}
)
\cup
\mathcal N_{j,r}^{(t)}
\right).
\end{align}
By
Lemma~\ref{lem:terminal_choice_path_independence},
\begin{equation}
\mathcal H_{j,r}^{(t)}
=
\operatorname{Ch}_j^{S,\star,(t)}
\left(
\mathcal A_{j,r-1}^{(t)}
\cup
\mathcal N_{j,r}^{(t)}
\right)
=
\operatorname{Ch}_j^{S,\star,(t)}
\left(
\mathcal A_{j,r}^{(t)}
\right).
\end{equation}
The statement therefore follows by induction. At the last invocation,
$\mathcal A_{j,r}^{(t)}
=
\mathcal A_j^{\star,(t)}$
and
$\mathcal H_{j,r}^{(t)}
=
\mathcal B_j^{\star,(t)}$,
which establishes
\eqref{eq:cumulative_terminal_choice}.
\end{proof}

\begin{lemma}[Persistence of terminal rejection]
\label{lem:terminal_rejection_persistence}
Under
Assumption~\ref{ass:terminal_choice_substitutability},
if a BE $b_i$ is presented to the terminal choice function of
$s_j$ and is rejected at some invocation, then it cannot belong to the
terminal retained set
$\mathcal B_j^{\star,(t)}$.
\end{lemma}

\begin{proof}
Suppose $b_i$ is rejected when the cumulative offer set is
$\mathcal A_{j,r}^{(t)}$, so that
$
b_i
\notin
\operatorname{Ch}_j^{S,\star,(t)}
\left(
\mathcal A_{j,r}^{(t)}
\right).
$
Since subsequent invocations only add newly explored BEs,
$
\mathcal A_{j,r}^{(t)}
\subseteq
\mathcal A_j^{\star,(t)}.
$
If, contrary to the claim,
$
b_i
\in
\operatorname{Ch}_j^{S,\star,(t)}
\left(
\mathcal A_j^{\star,(t)}
\right),
$
then substitutability applied to
$\mathcal A_{j,r}^{(t)}
\subseteq
\mathcal A_j^{\star,(t)}$
would imply
$
b_i
\in
\operatorname{Ch}_j^{S,\star,(t)}
\left(
\mathcal A_{j,r}^{(t)}
\right),
$
contradicting its earlier rejection.
Hence,
$
b_i
\notin
\operatorname{Ch}_j^{S,\star,(t)}
\left(
\mathcal A_j^{\star,(t)}
\right).
$
By
Lemma~\ref{lem:cumulative_terminal_offer_consistency},
the latter set equals
$\mathcal B_j^{\star,(t)}$.
\end{proof}

\begin{theorem}[Conditional restricted final-reapplication stability]
\label{thm:conditional_restricted_stability}
Suppose
Assumption~\ref{ass:terminal_choice_substitutability}
holds for every activated ES in the terminal explored
final-reapplication submarket. Then the terminal B\&E outcome admits
no restricted final-reapplication blocking pair in the sense of
Definition~\ref{def:restricted_final_blocking_pair}.
Moreover, within this explored submarket, every finally matched BE is
ex-ante proxy-rational, every activated ES obtains a
nonnegative updated proxy value from its terminal exact choice
(over the discretized feasible family), and the terminal contracts
satisfy the final contract-feasibility conditions established in
Theorem~\ref{thm:final_contract_feasibility}.
\end{theorem}

\begin{proof}
Final contract feasibility follows directly from
Theorem~\ref{thm:final_contract_feasibility}, while BE-side ex-ante
proxy rationality follows from
Lemma~\ref{lem:be_ex_ante_ir}.
For every activated ES, its terminal retained set is selected by the
terminal exact choice function under the terminal E\&C rate--price
pair. Since the empty set is feasible and has normalized updated proxy
value zero,
Proposition~\ref{prop:conditional_es_ex_ante_ir}
implies
\begin{equation}
\Pi_j^{S,\mathsf{upd},(t)}
\left(
\mathcal B_j^{\star,(t)}
\right)
\geq
0.
\label{eq:activated_es_terminal_ir}
\end{equation}
It remains to establish the absence of restricted blocking pairs.
Suppose, for contradiction, that
$(b_i,s_j)$
is a restricted final-reapplication blocking pair.
By
Definition~\ref{def:restricted_final_blocking_pair},
$
s_j
\in
\mathcal S_{\mathsf{act}}^{(t)}
\quad\text{and}\quad
b_i
\in
\mathcal A_j^{\star,(t)},
$
while
$
b_i
\notin
\mathcal B_j^{\star,(t)}.
$
By
Lemma~\ref{lem:cumulative_terminal_offer_consistency},
\begin{equation}
\mathcal B_j^{\star,(t)}
=
\operatorname{Ch}_j^{S,\star,(t)}
\left(
\mathcal A_j^{\star,(t)}
\right).
\label{eq:terminal_choice_in_stability_proof}
\end{equation}
Since
$b_i\in\mathcal A_j^{\star,(t)}$,
we have
$
\mathcal B_j^{\star,(t)}
\subseteq
\mathcal B_j^{\star,(t)}
\cup
\{b_i\}
\subseteq
\mathcal A_j^{\star,(t)}.
$
Applying the irrelevance-of-rejected-contracts property from
Lemma~\ref{lem:terminal_choice_irc}
to
\eqref{eq:terminal_choice_in_stability_proof}
therefore gives
\begin{equation}
\operatorname{Ch}_j^{S,\star,(t)}
\left(
\mathcal B_j^{\star,(t)}
\cup
\{b_i\}
\right)
=
\mathcal B_j^{\star,(t)}.
\label{eq:no_restricted_block_choice}
\end{equation}
Because
$b_i\notin\mathcal B_j^{\star,(t)}$,
\eqref{eq:no_restricted_block_choice}
implies
$
b_i
\notin
\operatorname{Ch}_j^{S,\star,(t)}
\left(
\mathcal B_j^{\star,(t)}
\cup
\{b_i\}
\right),
$
which contradicts
condition~\eqref{eq:restricted_block_es_choice}
in the definition of a restricted blocking pair.
Hence, no restricted final-reapplication blocking pair exists.
Together with the preceding feasibility and participation-rationality
properties, this establishes the stated conditional restricted
stability result.
\end{proof}

\begin{remark}[Scope of the restricted stability guarantee]
\label{rem:restricted_stability_scope}
Theorem~\ref{thm:conditional_restricted_stability}
does not establish global stability of the complete B\&E market.
First, the guarantee applies only to BE--ES pairs that are actually
presented to an ES-side terminal choice function in the explored
final-reapplication submarket. A pair that is never explored is not
covered.
Second, a B\&E contract released solely by the risk-restoration rule
before being presented to a terminal ES choice function is not treated
as an ES-choice rejection and is therefore outside the restricted
no-blocking guarantee.
Third, the theorem explicitly conditions on substitutability of the
terminal ES choice functions over the explored domains.
Substitutability is not asserted as an intrinsic property of the
general ES choice problem because heterogeneous reservation rates,
controlled-overbooking constraints, and aggregate shortfall-risk
coupling can generate complementarities across contracts.
Finally, the result concerns matching stability under the specified
ex-ante proxy preferences and terminal contract terms. It does not
imply incentive compatibility, strategy-proofness, truthfulness, or
global social-welfare optimality.
\end{remark}

\subsection{Online Fulfillment Feasibility and Efficiency}
\label{app:online_fulfillment_efficiency}

We finally characterize the feasibility and traffic-level efficiency
of the online fulfillment stage. Throughout this subsection, the
terminal B\&E contracts and E\&C supplementary reservations are fixed.
No rematching, price renegotiation, or additional cloud-side
reservation is allowed during online fulfillment.
For each ES $s_j$, define the aggregate contractual fulfillment demand
\begin{equation}
H_j^{(t)}
=
\sum_{b_i\in\mathcal F_j^{(t)}}
h_{i,j}^{(t)},
\label{eq:aggregate_contractual_demand}
\end{equation}
where
$
h_{i,j}^{(t)}
=
\min
\left\{
d_i^{B,(t)},
x_{i,j}^{B\leftrightarrow E,(t)}\Delta t
\right\}.
$
The local edge-side capacity available during the DPsT is
\begin{equation}
C_j^{S,(t)}
=
f_j^S\Delta t,
\label{eq:online_edge_capacity_volume}
\end{equation}
and the reserved cloud-side supplementary capacity available to
$s_j$ is
\begin{equation}
C_j^{C,(t)}
=
x_j^{E\leftrightarrow C,(t)}\Delta t.
\label{eq:online_cloud_capacity_volume}
\end{equation}

\begin{lemma}[Feasibility and work conservation of WPA]
\label{lem:wpa_feasibility_work_conservation}
Consider an invocation
$
\{a_i\}_{i\in\mathcal I}
=
\operatorname{WPA}
\left(
\{u_i\}_{i\in\mathcal I},
C,
\{\mathcal{O}_i\}_{i\in\mathcal I}
\right),
$
where $\mathcal I$ is finite,
$u_i\geq0$, and $C\geq0$.
The resulting allocation satisfies
\begin{equation}
0
\leq
a_i
\leq
u_i,
\qquad
\forall i\in\mathcal I,
\label{eq:wpa_individual_feasibility}
\end{equation}
and
\begin{equation}
\sum_{i\in\mathcal I}a_i
=
\min
\left\{
C,
\sum_{i\in\mathcal I}u_i
\right\}.
\label{eq:wpa_work_conservation}
\end{equation}
\end{lemma}

\begin{proof}
If
$
\sum_{i\in\mathcal I}u_i
\leq
C,
$
all demands can be fully satisfied, so
$a_i=u_i$
for every active demand and
\eqref{eq:wpa_individual_feasibility}--\eqref{eq:wpa_work_conservation}
hold immediately.
Now suppose
$
\sum_{i\in\mathcal I}u_i
>
C.
$
When the aggregate weighted demand is positive, WPA first distributes
the currently available capacity proportionally to the weighted demands.
If no provisional allocation exceeds the corresponding remaining
demand, the complete remaining capacity is assigned and the procedure
terminates.
If one or more provisional allocations exceed their corresponding
remaining demands, those allocations are truncated to the respective
demand levels. Hence, no BE receives more than its remaining demand.
The unused capacity created by truncation is then redistributed among
the still-unsatisfied BEs.
If the aggregate weight of the remaining BEs is zero, the prescribed
unweighted proportional fallback rule is applied. Therefore, the
allocation process cannot stall while both positive remaining demand
and positive available capacity exist.
By
Lemma~\ref{lem:online_fulfillment_termination},
the redistribution process terminates finitely. At termination, either
all demands have been satisfied or the available capacity has been
fully exhausted. Consequently,
$
\sum_{i\in\mathcal I}a_i
=
\min
\left\{
C,
\sum_{i\in\mathcal I}u_i
\right\}.
$
The truncation rule additionally guarantees
$0\leq a_i\leq u_i$
for every $i$.
\end{proof}

\begin{lemma}[Per-ES online fulfillment feasibility]
\label{lem:per_es_online_feasibility}
For every ES $s_j$ and every contracted BE
$b_i\in\mathcal F_j^{(t)}$, the total online contractual fulfillment
volume
\begin{equation}
g_{i,j}^{(t)}
=
y_{i,j}^{B\leftrightarrow E,(t)}
+
y_{i,j}^{E\leftrightarrow C,(t)}
\label{eq:total_contractual_fulfillment}
\end{equation}
satisfies
\begin{equation}
0
\leq
g_{i,j}^{(t)}
\leq
h_{i,j}^{(t)}.
\label{eq:per_be_contractual_fulfillment_feasibility}
\end{equation}
Moreover,
\begin{equation}
\sum_{b_i\in\mathcal F_j^{(t)}}
g_{i,j}^{(t)}
\leq
C_j^{S,(t)}
+
C_j^{C,(t)}.
\label{eq:per_es_total_capacity_feasibility}
\end{equation}
\end{lemma}

\begin{proof}
During edge-side fulfillment, WPA is applied to the contractual
demands
$\{h_{i,j}^{(t)}\}$.
By
Lemma~\ref{lem:wpa_feasibility_work_conservation},
\begin{equation}
0
\leq
y_{i,j}^{B\leftrightarrow E,(t)}
\leq
h_{i,j}^{(t)}.
\end{equation}
The remaining contractual demand is
$
d_{i,j}^{\mathsf{rem},(t)}
=
h_{i,j}^{(t)}
-
y_{i,j}^{B\leftrightarrow E,(t)}
\geq0.
$
Cloud-side WPA is subsequently applied only to these remaining demands.
Therefore,
\begin{equation}
0
\leq
y_{i,j}^{E\leftrightarrow C,(t)}
\leq
d_{i,j}^{\mathsf{rem},(t)}.
\end{equation}
Combining the two inequalities gives
$
0
\leq
y_{i,j}^{B\leftrightarrow E,(t)}
+
y_{i,j}^{E\leftrightarrow C,(t)}
\leq
h_{i,j}^{(t)},
$
which proves
\eqref{eq:per_be_contractual_fulfillment_feasibility}.
In addition, edge-side WPA never allocates more than
$C_j^{S,(t)}$, and cloud-side WPA never allocates more than
$C_j^{C,(t)}$. Hence,
\begin{align}
\sum_{b_i\in\mathcal F_j^{(t)}}
g_{i,j}^{(t)}
&=
\sum_{b_i\in\mathcal F_j^{(t)}}
y_{i,j}^{B\leftrightarrow E,(t)}
+
\sum_{b_i\in\mathcal F_j^{(t)}}
y_{i,j}^{E\leftrightarrow C,(t)}
\nonumber
\\
&\leq
C_j^{S,(t)}
+
C_j^{C,(t)}.
\end{align}
Thus, the online fulfillment outcome is physically feasible.
\end{proof}

\begin{lemma}[Per-ES work conservation]
\label{lem:per_es_online_work_conservation}
For every ES $s_j$, the aggregate contractual traffic fulfilled by the
edge-first and cloud-assisted online procedure satisfies
\begin{equation}
\sum_{b_i\in\mathcal F_j^{(t)}}
g_{i,j}^{(t)}
=
\min
\left\{
H_j^{(t)},
C_j^{S,(t)}+C_j^{C,(t)}
\right\}.
\label{eq:per_es_maximum_contractual_fulfillment}
\end{equation}
\end{lemma}

\begin{proof}
By
Lemma~\ref{lem:wpa_feasibility_work_conservation},
the total volume fulfilled locally at ES $s_j$ is
\begin{equation}
Y_j^{S,(t)}
=
\sum_{b_i\in\mathcal F_j^{(t)}}
y_{i,j}^{B\leftrightarrow E,(t)}
=
\min
\left\{
C_j^{S,(t)},
H_j^{(t)}
\right\}.
\label{eq:edge_total_work_conservation}
\end{equation}
Hence, the aggregate contractual demand remaining after edge-side
fulfillment is
\begin{align}
H_j^{\mathsf{rem},(t)}
&=
H_j^{(t)}
-
Y_j^{S,(t)}
\nonumber
\\
&=
\left[
H_j^{(t)}
-
C_j^{S,(t)}
\right]^+.
\label{eq:aggregate_remaining_contractual_demand}
\end{align}
Applying WPA to the cloud-side supplementary fulfillment gives
\begin{equation}
Y_j^{C,(t)}
=
\sum_{b_i\in\mathcal F_j^{(t)}}
y_{i,j}^{E\leftrightarrow C,(t)}
=
\min
\left\{
C_j^{C,(t)},
H_j^{\mathsf{rem},(t)}
\right\}.
\label{eq:cloud_total_work_conservation}
\end{equation}
Therefore,
\begin{align}
\sum_{b_i\in\mathcal F_j^{(t)}} g_{i,j}^{(t)}
&=
Y_j^{S,(t)} + Y_j^{C,(t)}
\nonumber\\
&=
\min\!\left\{ C_j^{S,(t)},\, H_j^{(t)} \right\}
\nonumber\\
&\quad+
\min\!\left\{ C_j^{C,(t)},\, H_j^{\mathsf{rem},(t)} \right\}
\nonumber\\
&=
\min\!\left\{ H_j^{(t)},\, C_j^{S,(t)} + C_j^{C,(t)} \right\}.
\end{align}
Thus, the online procedure fulfills the maximum aggregate contractual
traffic permitted by the fixed contractual demand and the edge-plus-cloud
capacity available to ES $s_j$.
\end{proof}

\begin{definition}[Feasible alternative contractual fulfillment]
\label{def:feasible_alternative_online_fulfillment}
For a fixed terminal contract outcome, an alternative contractual
fulfillment vector for ES $s_j$ is any collection
$
\left\{
\widetilde g_{i,j}^{(t)}
\right\}_{b_i\in\mathcal F_j^{(t)}}
$
satisfying
\begin{equation}
0
\leq
\widetilde g_{i,j}^{(t)}
\leq
h_{i,j}^{(t)},
\qquad
\forall b_i\in\mathcal F_j^{(t)},
\label{eq:alternative_individual_feasibility}
\end{equation}
and
\begin{equation}
\sum_{b_i\in\mathcal F_j^{(t)}}
\widetilde g_{i,j}^{(t)}
\leq
C_j^{S,(t)}
+
C_j^{C,(t)}.
\label{eq:alternative_capacity_feasibility}
\end{equation}
The terminal contracts and the ES-specific cloud reservation are held
fixed; supplementary capacity is not transferred across ESs.
\end{definition}

\begin{theorem}[Pareto efficiency of fulfilled contractual traffic]
\label{thm:online_contractual_traffic_pareto}
For any fixed target timeslot $t$, given the terminal B\&E contracts,
the retained E\&C supplementary reservations, and the realized
contractual demands, the online edge-first and cloud-assisted WPA
outcome is Pareto efficient with respect to fulfilled contractual
traffic.
Specifically, there exists no feasible alternative fulfillment
allocation
$\{\widetilde g_{i,j}^{(t)}\}$
such that
\begin{equation}
\widetilde g_{i,j}^{(t)}
\geq
g_{i,j}^{(t)}
\end{equation}
for every contracted BE, with strict inequality for at least one
contracted BE.
\end{theorem}

\begin{proof}
Suppose, for contradiction, that such a feasible alternative allocation
exists.
Then there is at least one contracted BE $b_{i^\star}$, associated
with some ES $s_{j^\star}$, such that
\begin{equation}
\widetilde g_{i^\star,j^\star}^{(t)}
>
g_{i^\star,j^\star}^{(t)},
\label{eq:strict_pareto_improvement}
\end{equation}
while
\begin{equation}
\widetilde g_{i,j^\star}^{(t)}
\geq
g_{i,j^\star}^{(t)}
\end{equation}
for every
$b_i\in\mathcal F_{j^\star}^{(t)}$.
Consequently,
\begin{equation}
\sum_{b_i\in\mathcal F_{j^\star}^{(t)}}
\widetilde g_{i,j^\star}^{(t)}
>
\sum_{b_i\in\mathcal F_{j^\star}^{(t)}}
g_{i,j^\star}^{(t)}.
\label{eq:strict_aggregate_improvement}
\end{equation}
By
Lemma~\ref{lem:per_es_online_work_conservation},
the WPA outcome already satisfies
\begin{equation}
\sum_{b_i\in\mathcal F_{j^\star}^{(t)}}
g_{i,j^\star}^{(t)}
=
\min
\left\{
H_{j^\star}^{(t)},
C_{j^\star}^{S,(t)}
+
C_{j^\star}^{C,(t)}
\right\}.
\label{eq:current_maximum_fulfillment}
\end{equation}
On the other hand, feasibility of the alternative allocation implies
both
\begin{equation}
\sum_{b_i\in\mathcal F_{j^\star}^{(t)}}
\widetilde g_{i,j^\star}^{(t)}
\leq
H_{j^\star}^{(t)}
\end{equation}
from the individual contractual-demand bounds and
\begin{equation}
\sum_{b_i\in\mathcal F_{j^\star}^{(t)}}
\widetilde g_{i,j^\star}^{(t)}
\leq
C_{j^\star}^{S,(t)}
+
C_{j^\star}^{C,(t)}
\end{equation}
from the available physical capacity.
Therefore,
\begin{equation}
\sum_{b_i\in\mathcal F_{j^\star}^{(t)}}
\widetilde g_{i,j^\star}^{(t)}
\leq
\min
\left\{
H_{j^\star}^{(t)},
C_{j^\star}^{S,(t)}
+
C_{j^\star}^{C,(t)}
\right\},
\end{equation}
which contradicts
\eqref{eq:strict_aggregate_improvement}
and
\eqref{eq:current_maximum_fulfillment}.
Hence, no feasible alternative allocation can weakly increase the
fulfilled contractual traffic of every contracted BE while strictly
increasing that of at least one contracted BE. The online fulfillment
outcome is therefore Pareto efficient with respect to fulfilled
contractual traffic.
\end{proof}

\begin{corollary}[Aggregate contractual-traffic maximality]
\label{cor:aggregate_contractual_traffic_maximality}
Under the same fixed terminal contracts and E\&C reservations, the
online fulfillment procedure maximizes the total fulfilled contractual
traffic:
\begin{equation}
\sum_{s_j\in\mathcal S}
\sum_{b_i\in\mathcal F_j^{(t)}}
g_{i,j}^{(t)}
=
\sum_{s_j\in\mathcal S}
\min
\left\{
H_j^{(t)},
C_j^{S,(t)}
+
C_j^{C,(t)}
\right\}.
\label{eq:global_contractual_traffic_maximality}
\end{equation}
\end{corollary}

\begin{proof}
The result follows by summing
\eqref{eq:per_es_maximum_contractual_fulfillment}
over all ESs. Since each BE is associated with at most one effective
B\&E contract and the retained cloud-side supplementary reservation is
ES-specific, the per-ES fulfillment problems are disjoint under the
fixed terminal contract outcome.
\end{proof}

\begin{remark}[Scope of the online efficiency guarantee]
\label{rem:online_efficiency_scope}
Theorem~\ref{thm:online_contractual_traffic_pareto}
is a traffic-level efficiency result under fixed terminal contracts and
fixed E\&C supplementary reservations.
It does not claim Pareto optimality with respect to the realized
participant utilities
$U_i^{B,(t)}$, $U_j^{S,(t)}$, or $U^{C,(t)}$,
because those utilities also contain reservation payments,
compensation transfers, operating costs, and uncovered-traffic losses.
The result also does not claim social-welfare optimality, optimality of
the offline contract-formation stage, or optimal reallocation of cloud
capacity across different ESs. In particular, the E\&C reservation of
each ES is held fixed during online fulfillment.
Finally, traffic exceeding the reserved B\&E volume is outside the
contractual fulfillment demand
$h_{i,j}^{(t)}$
and is therefore outside the Pareto comparison established above.
The WPA weights determine how scarce capacity is distributed among
competing contracted BEs, whereas the work-conservation property
ensures maximal aggregate fulfillment independently of the particular
positive weight values, with the prescribed unweighted fallback used
when the aggregate active weight is zero.
\end{remark}

\end{document}